\documentclass[12pt, twoside, epsf]{article}

\usepackage{color,graphicx,times}
\usepackage{amsmath, amssymb, graphics}

\newcommand{\mathsym}[1]{{}}
\newcommand{\unicode}[1]{{}}

\makeatletter
\long\def\M#1{\leavevmode\setbox\@tempboxa\hbox{#1}\@tempdima\fboxrule
    \advance\@tempdima \fboxsep \advance\@tempdima \dp\@tempboxa
   \hbox{\lower \@tempdima\hbox
  {\vbox{\hrule \@height \fboxrule
          \hbox{  \hskip\fboxsep
          \vbox{\vskip\fboxsep \box\@tempboxa\vskip\fboxsep}\hskip
                 \fboxsep\vrule \@width \fboxrule}%
                  }}}}
\makeatother

\let \ttorg \tt \def \tt{\ttorg \obeyspaces}

\begin{document}

\date{}

\title{\bf Iterants, Fermions and the Dirac Equation}

\author{Louis H. Kauffman \\
  Department of Mathematics, Statistics and Computer Science \\
  University of Illinois at Chicago \\
  851 South Morgan Street\\
  Chicago, IL, 60607-7045}

\maketitle
  
\thispagestyle{empty}

\section{Introduction}
 The simplest discrete system corresponds directly to the square root of minus one, when the square root of minus one is seen as an oscillation between plus and minus one. This way thinking about the square root of minus one as an {\em iterant} is explained below. More generally, by starting with a discrete time series of positions, one has immediately a non-commutativity of observations since the measurement of velocity involves the tick of the clock and the measurment of position does not demand the tick of the clock. Commutators that arise from discrete
observation generate a non-commutative calculus, and this calculus leads to a generalization of 
standard advanced calculus in terms of a non-commutative world. In a non-commutative world,
all derivatives are represented by commutators.  
\bigbreak

In this view, distinction and process arising from distinction is at the base of the world. Distinctions are elemental bits of awareness. The world is composed not of things but processes and observations.
We will discuss how basic Clifford algebra comes from very elementary processes like an alternation of $+-+-+-\cdots$  and the fact that one can think of $\sqrt{-1}$ itself as a temporal iterant, a product of an $\epsilon$ and an $\eta$ where the $\epsilon$ is the $+-+-+-\cdots$ and the $\eta$ is a time shift operator. Clifford algebra is at the base of the world! And the fermions are composed of these things.
\bigbreak

Secion 2 is an introduction to the process algebra of iterants and how the square root of minus one arises from an alternating process. Section 3 shows how iterants give an alternative way to do $2 \times 2$ matrix algebra. The section ends with the construction of the split quaternions. Section 4 considers iterants of arbitrary period (not just two) and shows, with the example of the cyclic group, how the ring of all $n \times n$ matrices can be seen as a faithful representation of an iterant algebra based on the cyclic group of order $n.$ We then generalize this construction to arbitrary non-commutative finite groups $G.$ Such a group has a multiplication table ($n \times n$  where $n$ is the order of the group $G.$). We show that by rearranging the multiplication table so the identity element appears on the diagonal, we get a set of permutation matrices that represent the group faithfully as $n \times n$ matrices. This gives a faithful representation of the iterant algebra associated with the group $G$ onto the ring of $n \times n$ matrices. As a result we see that iterant algebra is fundamental to all matrix algebra. Section 4 ends with a number of classical examples including iterant represtations for quaternion algebra. Section 5 goes back to $n \times n$ matrices and shows how the $2 \times 2$ iterant interpretation generalizes to an $n \times n$ matrix construction
 using the symmetric group $S_{n}.$ In Section 4 we have shown that there is a natural iterant algebra for $S_{n}$ that is associated with matrices of size $n! \times n!.$ In Section 5 we show there is another iterant algebra for $S_{n}$ associated with $n \times n$ matrices. We study this algebra and state some problems about its representation theory. 
Section 6 is a self-contained miniature version of the whole story in this paper, starting with 
the square root of minus one seen as a discrete oscillation, a clock. We proceed from there and analyze the position of the square root of minus one in relation to discrete systems and quantum mechanics.
We end this section by fitting together these observations into the structure of the Heisenberg commutator $$[p,q] = i \hbar.$$ Sections 6 and 7 show how iterants feature in discrete physics.
Section 8 discusses how Clifford algebras are fundamental to the structure of Fermions. We show how the simple algebra of the split quaternions, the very first iterant algebra that appears in relation to the square root of minus one, is in back of the structure of the operator algebra of the electron. The underlying Clifford structure describes a pair of Majorana Fermions, particles that are their own antiparticles. These Majorana Fermions can be symbolized by Clifford algebra generators $a$ and $b$ such that 
$a^2 = b^2 = 1$ and $ ab = -ba.$ One can take $a$ as the iterant corresponding to a period two oscillation, and $b$ as the time shifting operator. Then their product $ab$ is a square root of minus one in a non-commutative context. These are the Majorana Fermions that underlie an electron. The electron can be symbolized by $\phi = a + ib$ and the anti-electron by $\phi^{\dagger} = a - i b.$ These form the operator algebra for an electron. Note that 
$$\phi^2 = (a + ib)(a + ib) = a^2 - b^2 + i(ab + ba) = 0 + i0 = 0.$$
This nilpotent structure of the electron arises from its underlying Clifford structure in the form of a pair of Majorana Fermions. Section 8 then shows how braiding is related to the Majorana Femions. Section 9
discusses the fusion algebra for a Majorana Fermion in terms of the formal structure of the calculus of indications of G. Spencer-Brown \cite{LOF}. In this formalism we have a {\it logical particle} $P$ that is its own anti-particle. Thus $P$ interacts with itself to either produce itself or to cancel itself. Exactly such a formalism was devised by Spencer-Brown as a foundation for mathematics based on the concept of 
distinction. This section gives a short exposition of the calculus of indications and shows how, by way of iterants, the Fermion operators arise from recursive distinctions in the form of the re-entering mark. With this, we return to the square root of minus one in yet another way.
Section 10 discusses the structure of the Dirac equation and how the nilpotent and the Majorana operators arise naturally in this context. This section provides a link between our work and the work on nilpotent structures and the Dirac equation of Peter Rowlands \cite{Rowlands}. We end this section with an expression in split quaternions for the the Majorana Dirac equation in one dimension of time and three dimensions of space. The Majorana Dirac equation can be written as follows:
$$(\partial/\partial t + \hat{\eta} \eta \partial/\partial x + \epsilon \partial/\partial y + \hat{\epsilon} \eta \partial/\partial z - \hat{\epsilon} \hat{\eta} \eta m) \psi = 0$$ where $\eta$ and $\epsilon$ are the simplest generators of iterant algebra with $\eta^{2} = \epsilon^{2} = 1$ and $\eta \epsilon + \epsilon \eta = 0,$
and $\hat{\epsilon}, \hat{\eta}$ form a copy of this algebra that commutes with it. This combination of the simplest Clifford algebra with itself is the underlying structure of Majorana Fermions, forming indeed the underlying structure of all Fermions. The ending of the present paper forms the beginning of a study of the Majorana equation using iterants that will commence in sequels to this paper.
 \bigbreak
 
This paper is a stopping-place along the way in a larger story of processes, mathematics and physics that we are in the process of telling and exploring. To begin the story, we conclude this introduction with a fable about dice, time and the Schrodinger equation.
 \bigbreak

\subsection {\bf God Does Not Play Dice!}
Here is a little story about the square root of minus one and quantum mechanics.
\smallbreak

\noindent God said - I would really like to be able to base the universe on the Diffusion Equation
$$\partial \psi /\partial t = \kappa  \partial^{2} \psi /\partial x^{2}.$$ 
But I need to have some possibility for interference and waveforms.
And it should be simple. So I will just put a ``plus or minus" ambiguity into this equation, like so:
$$\pm \partial \psi /\partial t = \kappa \partial^{2} \psi /\partial x^{2}.$$ 
This is good, but it is not quite right. I do not play dice. The $\pm$ coefficient will have to be lawful, not random. Nothing is random. What to do? Aha! I shall take $\pm$ to mean the alternating sequence
$$\pm = \cdots + - + - + - + - \cdots$$ and time will become discrete. Then the equation will become a difference equation in space and time
$$\psi_{t+1} - \psi_{t} = (-1)^{t} \kappa  (\psi_{t}(x-dx) -2\psi_{t}(x) + \psi_{t}(x + dx))$$ where
$$ \partial_{x}^{2} \psi_{t} = \psi_{t}(x-dx) -2\psi_{t}(x) + \psi_{t}(x + dx).$$
This will do it, but I have to consider the continuum limit. But there is no meaning to 
$$(-1)^{t}$$ in the realm of continuous time. What do do? Ah! In the discrete world my wave function
(not a bad name for it!) divides into $\psi_{e}$ and $\psi_{o}$ where the time is either even or odd.
So I can write 
$$\partial_{t} \psi_{e} = \kappa \partial_{x}^{2} \psi_{o}$$  
$$\partial_{t} \psi_{o} = -\kappa \partial_{x}^{2} \psi_{e}.$$ 
I will take the continuum limit of $\psi_{e}$ and  $\psi_{o}$ {\it separately!}
 \bigbreak

Finally, a use for that so called imaginary number that 
Merlin has been bothering me with (You might wonder how Merlin could do this when I have not created him yet, but after all I am that am.). This $i$ has the property that $i^2 = -1$ so that 
$$i(A + iB) = iA - B$$ when $A$ and $B$ are ordinary numbers,
$$i = -1/i,$$ and so you see that if $i =1$ then $i = -1,$ and if $i=-1$ then $i=1.$ So $i$ just spends its time oscillating between $+1$ and $-1,$ but it does it lawfully and so I can regard it as a definition that 
$$i = \pm 1.$$ In fact, I can see now what Merlin what getting at. When I multiply $ii = (\pm1)(\pm1),$
I get $-1$ because the $i$ takes a little time to oscillate and so by the time this second term multiplies the first term, they are just out of phase and so we get either $(+1)(-1) = -1$ or $(-1)(+1) = -1.$ Either way,
$ii = -1$ and we have the perfect ambiguity. Heh. People will say that I am playing dice, but it is just 
not so.
Now $\pm 1$ behaves quite lawfully and I can write 
$$\psi = \psi_{e} + i \psi_{o}$$ so that 
$$i \partial_{t} \psi = i \partial_{t}( \psi_{e} + i\psi_{o})=  i \partial_{t}\psi_{e} -   \partial_{t}\psi_{o}$$
$$=i  \kappa \partial_{x}^{2} \psi_{o} + \kappa \partial_{x}^{2} \psi_{e} =  \kappa \partial_{x}^{2}( \psi_{e} + i\psi_{o})$$
$$ =\kappa \partial_{x}^{2} \psi.$$ 
Thus
$$i \partial \psi /\partial t = \kappa \partial^{2} \psi /\partial x^{2}.$$  

I shall call this the Schroedinger 
equation. Now I can rest on this seventh day before the real creation. This is the imaginary creation.
Instead of the simple diffusion equation, I have a mutual dependency where the temporal variation of 
$\psi_{e}$ is mediated by the spatial variation of $\psi_{o}$ and vice-versa. This is the price I pay for not playing dice.
$$\psi = \psi_{e} + i \psi_{o}$$
$$\partial_{t} \psi_{e} = \kappa \partial_{x}^{2} \psi_{o}$$  
$$\partial_{t} \psi_{o} = -\kappa \partial_{x}^{2} \psi_{e}.$$ 
$$i \partial \psi /\partial t = \kappa \partial^{2} \psi /\partial x^{2}.$$  

\noindent {\bf Remark.} The discrete recursion at the beginning of this tale, can actually be implemented to approximate solutions to the Schroedinger equation. This will be studied in a separate paper.
The reader may wish to point out that the playing of dice in quantum mechanics has nothing to do with the deterministic evolution of the Schroedinger equation, and everything to do with the measurment postulate that interprets $\psi \psi^{\dagger}$ as a probability density. The author (not God) agrees with the reader, but points out that God himself does not seem to have said anything about the measurement postulate. This postulate was born (or should we say Born?) after the Schoedinger equation was conceived. So we submit that it is not God who plays dice. 
\bigbreak

Probability and generalizations of classical probability are necessary for doing science. One should keep in mind that the quantum mechanics is based on a model that takes the solution of the Schroedinger equation to be a superposition of all possible observations of a given observer. The solution has norm equal to one in an appropriate vector space. That norm is the integral of the absolute square of the wave function over all of space. The absolute square of the wavefunction is seen as the associated probability density. This extraordinary and concise recipe for the probability of observed events is at the core of this subject. It is natural to ask, in relation to our fable, what is the relationship of probability for the diffusion process and the probability in quantum theory. This will have to be the subject of another paper and perhaps another fable.
\bigbreak

\noindent {\bf Acknowledgement.} It gives the author transfinite pleasure to thank G. Spencer-Brown, James Flagg, Alex Comfort, David Finkelstein, Pierre Noyes, Peter Rowlands, Sam Lomonaco and Bernd Schmeikal,  for conversations related to the considerations in this paper. Nothing here is their fault, yet Nothing would have happened without them. It gives the author further pleasure to thank the Mathematisches ForschungsInstitute Oberwolfach for its extraordinary hospitality during the final stages in writing this paper. 
\bigbreak

 \section{Iterants, Discrete Processes and Matrix Algebra}
The primitive idea behind an iterant is a periodic time series or 
``waveform" $$\cdots abababababab \cdots .$$ The elements of the waveform
can be any mathematically or empirically well-defined objects. We can regard
the ordered pairs $[a,b]$ and $[b,a]$ as abbreviations for the waveform or as two points
of view about the waveform ($a$ first or $b$ first). Call $[a,b]$ an {\em iterant}. 
One has the collection of
transformations of the form $T[a,b] = [ka, k^{-1}b]$ leaving the product $ab$ invariant.
This tiny model contains the seeds of special relativity, and the iterants contain the seeds of 
general matrix algebra! 
For related discussion see \cite{SS,SRF,SRCD,IML,KL,BL,Para,LOF}.
\bigbreak 

 Define products and sums of iterants as follows
 $$[a,b][c,d] = [ac,bd]$$  and $$[a,b] + [c,d] = [a+c,b+d].$$
 The operation of juxtapostion of waveforms is multiplication
 while $+$ denotes ordinary addition of ordered pairs. These operations are natural 
with respect to the 
 structural juxtaposition of iterants:
 $$...abababababab...$$
 $$...cdcdcdcdcdcd...$$ Structures combine at the points where they correspond.  
Waveforms combine at the times where they correspond. Iterants combine in juxtaposition.
\bigbreak

 If $\bullet$ denotes any form of binary compositon for the ingredients 
($a$,$b$,...) of
 iterants, then we can extend $\bullet$ to the iterants themselves by the 
definition
 $[a,b]\bullet[c,d] = [a\bullet c,b\bullet d]$.

 The appearance of a square root of minus
 one unfolds naturally from iterant considerations.  Define the ``shift" operator 
$\eta$ on iterants by the equation $$\eta[a,b] = [b,a]\eta$$ with $\eta^2 = 1.$  Sometimes it is 
convenient to think of
 $\eta$ as a delay opeator, since it shifts the waveform $...ababab...$  
by one internal
 time step. Now define  $$i = [-1,1]\eta $$
 We see at once that $$ii = [-1,1]\eta [-1,1]\eta = [-1,1] [1,-1] \eta^2 = [-1,1] [1,-1] = [-1,-1] = -1.$$  
Thus  $$ii=-1.$$     Here we have 
described $i$  in
 a {\em new} way as the superposition of the waveform 
 $\epsilon = [-1,1]$  and the temporal shift operator $\eta.$
By writing $i = \epsilon \eta$ we recognize an active version of the waveform that shifts temporally when it is observed.
This theme of including the result of time in observations of a discrete system occurs at the foundation of our construction.
\bigbreak

In the next section we show how all of matrix algebra can be formulated in terms of iterants.
\bigbreak

 \section{MATRIX ALGEBRA VIA ITERANTS}
Matrix algebra has some strange wisdom built into its very bones.
Consider a two dimensional periodic pattern or ``waveform."
$$......................$$
 $$...abababababababab...$$
 $$...cdcdcdcdcdcdcdcd...$$
 $$...abababababababab...$$
 $$...cdcdcdcdcdcdcdcd...$$
 $$...abababababababab...$$
 $$......................$$
 
$$\left(\begin{array}{cc}
a&b\\
c&d
\end{array}\right), \left(\begin{array}{cc}
b&a\\
d&c
\end{array}\right), \left(\begin{array}{cc}
c&d\\
a&b
\end{array}\right), \left(\begin{array}{cc}
d&c\\
b&a
\end{array}\right)$$ Above are some of the matrices apparent in this array.
\noindent Compare the matrix with the ``two dimensional waveform" shown 
above. A given matrix freezes out a way to view the infinite waveform.
In order to keep track of this patterning, lets write
 
 $$[a,b] + [c,d]\eta = 	\left(\begin{array}{cc}
			a&c\\
			d&b
			\end{array}\right).$$ 

\noindent where
$$[x,y] = 	\left(\begin{array}{cc}
			x&0\\
			0&y
			\end{array}\right).$$ 

\noindent and
$$\eta = 	\left(\begin{array}{cc}
			0&1\\
			1&0
			\end{array}\right).$$ 

Recall the definition of matrix multiplication.
$$\left(\begin{array}{cc}
			a&c\\
			d&b
			\end{array}\right)
			\left(\begin{array}{cc}
			e&g\\
			h&f
			\end{array}\right) =
			\left(\begin{array}{cc}
			ae+ch&ag+cf\\
			de+bh&dg+bf
			\end{array}\right).
			$$ 
			Compare this with the iterant multiplication.
			$$([a,b] + [c,d]\eta)([e,f]+[g,h]\eta) = $$
			$$[a,b][e,f] + [c,d]\eta[g,h]\eta + [a,b][g,h]\eta + [c,d]\eta[e,f] =$$
			$$[ae,bf] + [c,d][h,g] +( [ag, bh] + [c,d][f,e])\eta = $$
			$$[ae,bf] +[ch,dg] + ( [ag, bh] + [cf,de])\eta = $$
			$$[ae+ch, dg+bf] + [ag + cf, de+bh]\eta.$$
Thus matrix multiplication is identical with iterant multiplication. The concept of the iterant can be used to
motivate matrix multiplication.
\bigbreak

\noindent The four matrices that can be framed in the two-dimensional wave 
form are all obtained from the two iterants
 $[a,d]$ and $[b,c]$ via the shift operation $\eta [x,y] = [y,x] \eta$ which we 
shall denote by  an overbar as shown below  $$\overline{[x,y]} = [y,x].$$
  
\noindent Letting  $A = [a,d]$  and $B=[b,c]$, we see that the four matrices seen in the 
grid are $$A + B \eta, B + A \eta, \overline{B} + \overline{A}\eta,
  \overline{A} + \overline{B}\eta.$$

 \noindent The operator  $\eta$  has the effect of rotating an iterant by ninety 
 degrees in the formal plane. Ordinary matrix multiplication can be written in a
 concise form using the following rules:

 $$\eta \eta = 1$$
 $$\eta Q = \overline{Q} \eta$$  where  Q is any two element iterant.
 Note the correspondence
 $$                      \left(\begin{array}{cc}
			a&b\\
			c&d
			\end{array}\right)
			=
			\left(\begin{array}{cc}
			a&0\\
			0&d
			\end{array}\right) 
			\left(\begin{array}{cc}
			1&0\\
			0&1
			\end{array}\right)
		        +
		        \left(\begin{array}{cc}
			b&0\\
			0&c
			\end{array}\right)
			\left(\begin{array}{cc}
			0&1\\
			1&0
			\end{array}\right) 
			= [a,d]1 + [b,c]\eta.$$ 
This means that $[a,d]$ corresponds to a diagonal matrix.
			 $$[a,d] = 	\left(\begin{array}{cc}
			a&0\\
			0&d
			\end{array}\right),$$ 
			$\eta$ corresponds to the anti-diagonal permutation matrix.
 $$\eta = 	\left(\begin{array}{cc}
			0&1\\
			1&0
			\end{array}\right),$$ 
			and $[b,c]\eta$ corresponds to the product of a diagonal matrix and the permutation matrix. $$[b,c]\eta = 	\left(\begin{array}{cc}
			b&0\\
			0&c
			\end{array}\right)
			\left(\begin{array}{cc}
			0&1\\
			1&0
			\end{array}\right) =
			 \left(\begin{array}{cc}
			0&b\\
			c&0
			\end{array}\right).$$  
			Note also that 		
$$\eta [c,b] = 
\left(\begin{array}{cc}
			0&1\\
			1&0
			\end{array}\right)
				 \left(\begin{array}{cc}
			c &0\\
			0&b
			\end{array}\right) =
			 \left(\begin{array}{cc}
			0&b\\
			c&0
			\end{array}\right).$$
			This is the matrix interpretation of the equation
			$$[b,c]\eta = \eta [c,b].$$
			\bigbreak
			
The fact that the iterant expression $ [a,d]1 + [b,c]\eta$ captures the whole of $2 \times 2$ matrix algebra corresponds to the fact that a two by two matrix is
combinatorially the union of the identity pattern (the diagonal)  and the interchange pattern (the antidiagonal) that correspond to the operators $1$ and $\eta.$
$$\left(\begin{array}{cc}
			*&@\\
			@ & *\\
			\end{array}\right)$$
			In the formal diagram for a matrix shown above, we indicate the diagonal by $*$ and the anti-diagonal by $@.$
\bigbreak

In the case of complex numbers we represent 
$$\left(\begin{array}{cc}
			a&-b\\
			b&a
			\end{array}\right) = [a,a] + [-b,b]\eta = a1 + b[-1,1]\eta = a + bi.$$ 
			In this way, we see that all of $2 \times 2$ matrix algebra is a hypercomplex number system based on the symmetric group $S_{2}.$
			In the next section we generalize this point of view to arbirary finite groups.
			\bigbreak 
 
 \noindent We have reconstructed the square root of minus one in the form of 
the matrix
   $$ i = \epsilon \eta = [-1,1]\eta 
   =\left(\begin{array}{cc}
			0&-1\\
			1&0
			\end{array}\right).$$
In this way, we arrive at this well-known representation of the complex numbers in terms of matrices.
Note that if we identify  the ordered pair $(a,b)$ with $a +ib,$ then this means taking the identification
$$(a,b) =  \left(\begin{array}{cc}
			a&-b\\
			b&a
			\end{array}\right).$$ Thus the geometric interpretation of multiplication by $i$ as a ninety degree rotation in the Cartesian plane, 
			 $$i(a,b) = (-b,a),$$ takes the place of the matrix equation
$$ i (a,b) = \left(\begin{array}{cc}
			0&-1\\
			1&0
			\end{array}\right)
			 \left(\begin{array}{cc}
			a&-b\\
			b&a
			\end{array}\right)
			=  \left(\begin{array}{cc}
			-b&-a\\
			a&-b
			\end{array}\right) =b + ia = (-b,a).$$
In iterant terms we have $$i[a,b] = \epsilon \eta [a,b] = [-1,1] [b,a] \eta  = [-b,a] \eta,$$ and this corresponds to the matrix equation
$$ i [a,b] = \left(\begin{array}{cc}
			0&-1\\
			1&0
			\end{array}\right)
			 \left(\begin{array}{cc}
			a& 0\\
			0 &b
			\end{array}\right)
			=  \left(\begin{array}{cc}
			0&-b\\
			a& 0
			\end{array}\right) =[-b,a] \eta.$$ All of this points out how the complex numbers, as we have previously examined them, live naturally in the context of the non-commutative algebras of iterants and  matrices. The factorization of $i$ into a product $\epsilon \eta$ of non-commuting iterant operators is closer both to the temporal nature of $i$ and to its algebraic roots.
\bigbreak
			
\noindent  More generally, we see that 
 $$(A + B\eta)(C+D\eta) = (AC+B\overline{D}) + (AD + 
B\overline{C})\eta$$

 \noindent writing the $2 \times 2$ matrix algebra as a system of 
 hypercomplex numbers.   Note that 
 $$(A+B\eta)(\overline{A}-B\eta) = A\overline{A} - B\overline{B}$$
 
 \noindent The formula on the right equals the determinant of the 
 matrix. Thus we define the {\em conjugate} of
 $Z = A+B\eta$ by the formula 
$$\overline{Z} = \overline{A+B\eta} = \overline{A} - B\eta,$$ and we have the formula
$$D(Z) = Z \overline{Z}$$ for the determinant $D(Z)$ where
$$Z= A + B\eta = 
\left(\begin{array}{cc}
			a&c\\
			d&b
			\end{array}\right)$$ where $A=[a,b]$ and $B=[c,d].$ Note
			that $$A\overline{A} =[ab, ba] = ab1 = ab,$$ so that
			$$D(Z) = ab -cd.$$ Note also that we assume that $a,b,c,d$ are in a commutative base ring. 
	\bigbreak	
	Note also that for $Z$ as above, 
	$$\overline{Z} = \overline{A} - B\eta =
	\left(\begin{array}{cc}
			b&-c\\
			-d&a
			\end{array}\right).$$ This is the classical adjoint of the matrix $Z.$
			\bigbreak
		
			We leave it to the reader to check that for matrix iterants $Z$ and $W,$
			$$Z\overline{Z} = \overline{Z}Z$$ and that 
			$$\overline{ZW} = \overline{W}\overline{Z}$$ and 
			$$\overline{Z + W} = \overline{Z} + \overline{W}.$$
			Note also that $$\overline{\eta} = - \eta, $$ whence
			$$\overline{B\eta} = - B\eta = -\eta \overline{B} = \overline{\eta} \overline{B}.$$
			We can prove that 
			$$D(ZW) = D(Z)D(W)$$ as follows
			$$D(ZW) = ZW \overline{ZW} = ZW \overline{W} \,\overline{Z} =
			 Z \overline{Z}W \overline{W} = D(Z)D(W).$$
			 Here the fact that $W \overline{W}$ is in the base ring which is commutative allows us to remove it from in between the appearance of $Z$ and $\overline{Z}.$ Thus we see that 
			 iterants as $2 \times 2$ matrices form a direct non-commutative generalization of 
			 the complex numbers.
 \bigbreak

 It is worth pointing out the first precursor to the quaternions ( the so-called {\it split quaternions}): This 
 precursor is the system  $$\{\pm{1}, \pm{\epsilon}, \pm{\eta}, 
\pm{i}\}.$$
 Here $\epsilon\epsilon = 1 = \eta\eta$ while $i=\epsilon \eta$ so 
that $ii = -1$.  
 The basic operations in this
 algebra are those of epsilon and eta.  Eta is the delay shift operator 
that reverses
 the components of the iterant. Epsilon negates one of the 
components, and leaves the
 order unchanged. The quaternions arise directly from these two 
operations once
 we construct an extra
 square root of minus one that commutes with them. Call this extra
 root of minus one $\sqrt{-1}$. Then the quaternions are generated 
by 
 $$I=\sqrt{-1}\epsilon, J= \epsilon \eta, K= \sqrt{-1}\eta$$
  with $$I^{2} = J^{2}=K^{2}=IJK=-1.$$
 The ``right" way to generate the quaternions is to start at the bottom 
iterant level
 with boolean values of $0$ and $1$ and the operation EXOR (exclusive or). Build 
iterants on this,
 and matrix algebra from these iterants.  This gives the square root 
of negation. Now
 take pairs of values from this new algebra and build $2 \times 2$ matrices 
again.  
 The coefficients include square roots of negation that commute with 
constructions at the
next level and so quaternions appear in the third level of this 
hierarchy. We will return to the quaternions after discussing other examples that involve matrices of all sizes.
\bigbreak

\section {Iterants of Arbirtarily High Period}
As a next example, consider a waveform of period three.
$$\cdots abcabcabcabcabcabc \cdots$$
Here we see three natural iterant views (depending upon whether one starts at $a$, $b$ or $c$).
$$[a,b,c],\,\,\, [b,c,a], \,\,\, [c,a,b].$$ The appropriate shift operator is given by the formula
$$[x,y,z]S = S[z,x,y].$$ Thus, with $T = S^{2},$
$$[x,y,z]T = T[y,z,x]$$ and $S^{3} = 1.$
With this we obtain a closed algebra of iterants whose general element is of the form $$[a,b,c] + [d,e,f]S + [g,h,k]S^{2}$$ where $a,b,c,d,e,f,g,h,k$ are real or complex numbers.
Call this algebra $\mathbb{V}ect_{3}(\mathbb{R})$ when the scalars are in a commutative ring with unit  $\mathbb{F}.$ Let $M_{3}(\mathbb{F})$ denote the $3 \times 3$ matrix algebra over $\mathbb{F}.$ We have the 
\smallbreak 

\noindent {\bf Lemma.} The iterant algebra  $\mathbb{V}ect_{3}(\mathbb{F})$ is isomorphic to the full
$3 \times 3$ matrix algebra $M_{3}((\mathbb{F}).$
\smallbreak

\noindent {\bf Proof.} Map $1$ to the matrix
$$\left(\begin{array}{ccc}
			1&0&0\\
			0&1&0\\
			0&0&1
			\end{array}\right).$$
Map $S$ to the matrix 
$$\left(\begin{array}{ccc}
			0&1&0\\
			0&0&1\\
			1&0&0
			\end{array}\right),$$
                          and map $S^2$ to the matrix
                           $$\left(\begin{array}{ccc}
			0&0&1\\
			1&0&0\\
			0&1&0
			\end{array}\right),$$
			Map $[x,y,z]$ to the diagonal matrix
			$$\left(\begin{array}{ccc}
			x&0&0\\
			0&y&0\\
			0&0&z
			\end{array}\right).$$
			Then it follows that $$[a,b,c] + [d,e,f]S + [g,h,k]S^{2}$$ maps to the matrix
			$$\left(\begin{array}{ccc}
			a&d&g\\
			h&b&e\\
			f&k&c
			\end{array}\right),$$ preserving the algebra structure. Since any $3 \times 3$ matrix can be written uniquely in this form, 
			it follows that  $\mathbb{V}ect_{3}(\mathbb{F})$ is isomorphic to the full
                          $3 \times 3$ matrix algebra $M_{3}(\mathbb{F}).$ $//$

We can summarize the pattern behind this expression of $3 \times 3$ matrices  by the following symbolic matrix.
$$\left(\begin{array}{ccc}
			1&S&T\\
			T&1&S\\
			S&T&1
			\end{array}\right)$$
			Here the letter $T$ occupies the positions in the matrix that correspond to the permutation matrix that represents it, and the letter $T = S^2$ occupies
			the positions corresponding to its permutation matrix.  The $1$'s occupy the diagonal for the corresponding identity matrix. The iterant representation
			corresponds to writing the $3 \times 3$ matrix as a disjoint sum of these permutation matrices such that the matrices themselves are closed under multiplication.
			In this case the matrices form a permutation representation of the cyclic group of order $3$, $C_{3} = \{1, S, S^{2} \}.$
			\bigbreak
			
\noindent {\bf Remark.} Note that a permutation matrix is a matrix of zeroes and ones such that some 
permutation of the rows of the matrix transforms it to the identity matrix. Given an $n \times n$ permutation matrix $P,$ we associate to it a permuation 
$$\sigma(P):\{1,2, \cdots, n\} \longrightarrow \{1,2, \cdots, n\}$$ via the following formula
$$i \sigma(P) = j$$ where $j$ denotes the column in $P$ where the $i$-th row has a $1$.
Note that an element of the  domain of a permutation is indicated to the left of the symbol for the permutation.
It is then easy to check that for permutation matrices $P$ and $Q$, 
$$\sigma(P)\sigma(Q) = \sigma(PQ)$$ given that we compose the permutations from left to right according to this convention.
\bigbreak

			It should be clear to the reader that this construction generalizes directly for iterants of any period and hence for a set of operators forming a cyclic group of any order.
			In fact we shall generalize further to any finite 
			group $G.$ We now define $\mathbb{V}ect_{n(}G,\mathbb{F})$ for any finite group $G.$
			\bigbreak
			
			\noindent {\bf Definition.} Let $G$ be a finite group, written multiplicatively. Let $\mathbb{F}$ denote a given commutative ring with unit. Assume that $G$ acts as a group of permutations on the set $\{1,2, 3,\cdots, n \}$ so that given an element $g \in G$ we have (by abuse of notation) $$g: \{1,2, 3,\cdots, n \} \longrightarrow \{1,2, 3,\cdots, n \}.$$
			We shall write $$ig$$ for the image of $i \in \{1,2, 3,\cdots, n \}$ under the permutation represented by $g.$ Note that this denotes functionality from the left and so we ask that $(ig)h = i(gh)$ for all elements $g, h \in G$ and $i1 = i$ for all $i$, in order to have a representation of $G$ as permutations. We shall call an $n$-tuple of elements of 
			$\mathbb{F}$ a {\it vector } and denote it by $a = (a_{1},a_{2},\cdots, a_{n}).$ We then define an action of $G$ on vectors over $\mathbb{F}$ by the formula
			$$a^{g} = (a_{1g}, a_{2g}, \cdots, a_{ng}),$$ and note that $(a^{g})^{h} = a^{gh}$ for all $g,h \in G.$ We now define an algebra  $\mathbb{V}ect_{n}(G,\mathbb{F})$, the 
			{\it iterant algebra for $G$,} to be the set of finite sums of formal products of vectors and group elements in the form $ag$ with multiplication rule
			$$(ag)(bh) = ab^{g}(gh),$$ and the understanding that $(a + b)g = ag + bg$ and for all vectors $a,b$ and group elements $g.$ It is understood that vectors are added 
			coordinatewise and multiplied coordinatewise. Thus $(a + b)_{i} = a_{i} + b_{i}$  and $(ab)_{i} = a_{i}b_{i}.$
			\bigbreak
			
			\noindent {\bf Theorem.} Let G be a finite group of order $n.$ Let $\rho: G \longrightarrow S_{n}$ denote the right regular representation of $G$ as permutations of
			$n$ things where we list the elements of $G$ as $G = \{ g_{1}, \cdots , g_{n}\}$ and let $G$ act on its own underlying set via the definition
			 $ g_{i} \rho(g) = g_{i}g.$  Here we describe $\rho(g)$ acting on the set of elements $g_{k}$ of $G.$ If we wish 
			 to regard $\rho(g)$ as a mapping of the set $\{1,2,\cdots n\}$ then we replace $g_{k}$ by $k$ and 
			$ i\rho(g) = k$ where $g_{i}g = g_{k}.$
			\smallbreak
			 Then $\mathbb{V}ect_{n}(G,\mathbb{F})$ is isomorphic to the matrix algebra $M_{n}((\mathbb{F}).$ In particular, we have that
			 $\mathbb{V}ect_{n!}(S_{n},\mathbb{F})$ is isomorphic with the matrices of size $n! \times n!$, $M_{n!}((\mathbb{F}).$
			 \smallbreak
			 
			 \noindent {\bf Proof.} Consider the $n \times n$ matrix consisting in the multiplication table for $G$ with the columns and rows listed in the order
			  $[ g_{1}, \cdots , g_{n}].$ Permute the rows of this table so that the diagonal consists in all $1$'s. Let the resulting table be called the $G$-{\it Table}. 
			  The  $G$-{\it Table} is labeled by elements of the group. For a vector $a,$ let $D(a)$ denote the $n \times n$ diagonal matrix whose entries in order down the diagonal 
			  are the entries of $a$ in the order specified by $a.$
			  For each group element $g$, let $P_{g}$ denote the permutation matrix with $1$ in every spot on the  $G$-{\it Table} that is labeled by $g$ and $0$ in all other spots.
			  It is now a direct verification that the mapping $$F(\Sigma_{i=1}^{n} a_{i}g_{i}) = \Sigma_{i=1}^{n} D(a_{i})P_{g_{i}}$$  defines an isomorphism from 
			  $\mathbb{V}ect_{n}(G,\mathbb{F})$ to the matrix algebra $M_{n}((\mathbb{F}).$ The main point to check is that 
			  $\sigma(P_{g}) = \rho(g).$ We now prove this fact.
			  \bigbreak
			  
			  \noindent In the  $G$-{\it Table} the rows correspond to 
			  $$\{g_{1}^{-1},g_{2}^{-1},\cdots g_{n}^{-1}\}$$ and the columns correspond to 
			  $$\{g_{1},g_{2},\cdots g_{n}\}$$ so that the $i$-$i$ entry of the table is
			  $g_{i}^{-1} g_{i} = 1.$ With this we have that in the table, a group element $g$ occurs in the $i$-th row at column $j$ where
			  $$g_{i}^{-1}g_{j} = g.$$ This is equivalent to the equation
			  $$g_{i}g = g_{j}$$ which, in turn is equivalent to the statement
			  $$i \rho(g) = j.$$ This is exactly our functional interpretation of the action of the permutation 
			  corresponding to the matrix $P_{g}.$ Thus $$\rho(g) = \sigma(P_{g}).$$ The remaining detalls of the proof are straightforward and left to the reader.
			  $//$
			  \bigbreak

\noindent {\bf Examples.} 
\begin{enumerate}
\item We have already implicitly given examples of this process of translation.
Consider the cyclic group of order three. $$C_{3} = \{1,S,S^2 \}$$ with $S^3 = 1.$ The multiplication table is 
$$\left(\begin{array}{ccc}
			1&S&S^2\\
			S&S^2&1\\
			S^2&1&S
			\end{array}\right).$$
			Interchanging the second and third rows, we obtain
			$$\left(\begin{array}{ccc}
			1&S&S^2\\
			S^2&1&S\\
			S&S^2&1
			\end{array}\right),$$ and this is the  $G$-{\it Table} that we used for 
			$\mathbb{V}ect_{3}(C_{3},\mathbb{F})$ prior to proving the Main Theorem.
			\smallbreak
			
			The same pattern works for abitrary cyclic groups. for example, consider the cyclic group of order $6.$ $C_{6} = \{1,S,S^2,S^3,S^4,S^5\}$ with $S^6 = 1.$ The multiplication table is
			$$\left(\begin{array}{cccccc}
			1&S&S^2&S^3&S^4&S^5\\
			S&S^2&S^3&S^4&S^5&1\\
			S^2&S^3&S^4&S^5&1&S\\
			S^3&S^4&S^5&1&S&S^2\\
			S^4&S^5&1&S&S^2&S^3\\
                            S^5&1&S&S^2&S^3&S^4\\
			\end{array}\right).$$
Rearranging to form the  $G$-{\it Table}, we have
$$\left(\begin{array}{cccccc}
			1&S&S^2&S^3&S^4&S^5\\
			 S^5&1&S&S^2&S^3&S^4\\
			 S^4&S^5&1&S&S^2&S^3\\
			 S^3&S^4&S^5&1&S&S^2\\
	                    S^2&S^3&S^4&S^5&1&S\\
			S&S^2&S^3&S^4&S^5&1\\
			\end{array}\right).$$
The permutation matrices corresponding to the positions of $S^{k}$ in the  $G$-{\it Table} give the matrix representation that gives the isomorphsm of $\mathbb{V}ect_{6}(C_{6},\mathbb{F})$ with the full algebra of 
six by six matrices.			
\item Now consider the symmetric group on six letters, $$S_{6}= \{1,R,R^2,F,RF,R^{2}F \}$$ where $R^{3}=1, F^{2}=1, FR = RF^{2}.$ Then the multiplication table is
$$\left(\begin{array}{cccccc}
			1&R&R^2&F&RF&R^{2}F\\
			 R&R^{2}&1&RF&R^{2}F&F\\
			 R^{2}&1&R&R^{2}F&F&RF\\
			 F&R^{2}F&RF&1&R^2&R\\
			 RF&F&R^{2}F&R&1&R^{2}\\
			  R^{2}F&RF&F&R^{2}&R&1\\
			\end{array}\right).$$
			The corresponnding  $G$-{\it Table} is
			$$\left(\begin{array}{cccccc}
			1&R&R^2&F&RF&R^{2}F\\
			 R^{2}&1&R&R^{2}F&F&RF\\
			  R&R^{2}&1&RF&R^{2}F&F\\
			 F&R^{2}F&RF&1&R^2&R\\
			 RF&F&R^{2}F&R&1&R^{2}\\
			  R^{2}F&RF&F&R^{2}&R&1\\
			\end{array}\right).$$
			Here is a rewritten version of the  $G$-{\it Table} with 
			$$R = \Delta, R^2 = \Theta, F = \Psi, RF = \Omega, R^{2}F = \Sigma.$$
$$\left(\begin{array}{cccccc}
			1& \Delta&\Theta&\Psi& \Omega&\Sigma\\
			 \Theta&1& \Delta&\Sigma&\Psi& \Omega\\
			   \Delta&\Theta&1& \Omega&\Sigma&\Psi\\
			 \Psi&\Sigma& \Omega&1& \Theta&\Delta\\
			  \Omega&\Psi&\Sigma& \Delta&1&\Theta\\
			  \Sigma& \Omega&\Psi&\Theta& \Delta&1\\
			\end{array}\right).$$
			This  $G$-{\it Table} is the keystone for the isomorphism of $\mathbb{V}ect_{6}(S_{3},\mathbb{F})$ with the full algebra of six by six matrices. At this point it may occur to the reader to wonder about 
 $\mathbb{V}ect_{3}(S_{3},\mathbb{F})$ since $S_{3}$ does act on vectors of length three. We will discuss
  $\mathbb{V}ect_{n}(S_{n},\mathbb{F})$	in the next section. We see from this example how it will come about that
  $\mathbb{V}ect_{n!}(S_{n},\mathbb{F})$ is isomorphic with the full algebra of $n! \times n!$ matrices.	In particular, here are the permutation matrices that form the non-identity elements of this representation of
  the symmetric group on three letters.
  $$ R = \Delta = \left(\begin{array}{cccccc}
			0&1&0&0&0&0\\
			 0&0&1&0&0&0\\
			  1&0&0&0&0&0\\
			 0&0&0&0&0&1\\
			  0&0&0&1&0&0\\
			  0&0&0&0&1&0\\
			\end{array}\right)$$
$$ R^2 = \Theta = \left(\begin{array}{cccccc}
			0&0&1&0&0&0\\
			 1&0&0&0&0&0\\
			   0&1&0&0&0&0\\
			 0&0&0&0&1&0\\
			  0&0&0&0&0&1\\
			  0&0&0&1&0&0\\
			\end{array}\right)$$
$$ F = \Psi = \left(\begin{array}{cccccc}
			0&0&0&1&0&0\\
			 0&0&0&0&1&0\\
			  0&0&0&0&0&1\\
			 1&0&0&0&0&0\\
			  0&1&0&0&0&0\\
			  0&0&1&0&0&0\\
			\end{array}\right)$$
$$ FR = \Omega = \left(\begin{array}{cccccc}
			0&0&0&0&1&0\\
			 0&0&0&0&0&1\\
			  0&0&0&1&0&0\\
			 0&0&1&0&0&0\\
			  1&0&0&0&0&0\\
			  0&1&0&0&0&0\\
			\end{array}\right)$$
$$ FR^2 = \Sigma  = \left(\begin{array}{cccccc}
			0&0&0&0&0&1\\
			 0&0&0&1&0&0\\
			  0&0&0&0&1&0\\
			 0&1&0&0&0&0\\
			  0&0&1&0&0&0\\
			  1&0&0&0&0&0\\
			\end{array}\right)$$

\item In this example we consider the group $G = C_{2} \times C_{2},$ often called the ``Klein $4$-Group." We take $G = \{1,A,B,C\}$ where $A^2 = B^2 = C^2 = 1, AB = BA = C.$ Thus $G$ has the multiplication table, which is also its  $G$-{\it Table} for $\mathbb{V}ect_{4}(G,\mathbb{F}).$
$$\left(\begin{array}{cccc}
			1& A&B&C\\
			 A&1&C&B\\
			  B&C&1& A\\
			 C&B&A&1\\
			\end{array}\right).$$
			Thus we have the following permutation matrices that I shall call $E, A, B, C:$
			$$E = \left(\begin{array}{cccc}
			1& 0&0&0\\
			 0&1&0&0\\
			  0&0&1& 0\\
			 0&0&0&1\\
			\end{array}\right),$$
			$$A = \left(\begin{array}{cccc}
			0&1&0&0\\
			 1&0&0&0\\
			 0&0&0&1\\
			 0&0&1&0\\
			\end{array}\right),$$
			$$B = \left(\begin{array}{cccc}
			0&0&1&0\\
			 0&0&0&1\\
			 1&0&0&0\\
			 0&1&0&0\\
			\end{array}\right),$$
			$$C = \left(\begin{array}{cccc}
			0&0&0&1\\
			 0&0&1&0\\
			 0&1&0&0\\
			 1&0&0&0\\
			\end{array}\right).$$ The reader will have no difficulty verifying that 
			 $A^2 = B^2 = C^2 = 1, AB = BA = C.$
			 Recall that $[x,y,z,w]$ is iterant notation for the diagonal matrix
			 $$[x,y,z,w] = \left(\begin{array}{cccc}
			x&0&0&1\\
			 0&y&1&0\\
			 0&1&z&0\\
			 1&0&0&w\\
			\end{array}\right).$$ Let 
			$$\alpha = [1,-1,-1,1], \beta = [1,1,-1,-1], \gamma = [1,-1,1,-1].$$
			And let $$I = \alpha A, J = \beta B, K = \gamma C.$$
			Then the reader will have no trouble verifying that 
			$$I^2 = J^2 = K^2 = IJK = -1, IJ = K, JI = -K.$$
			Thus we have constructed the quaternions as iterants in relation to the Klein Four Group.
			in Figure~\ref{fourgroup} we illustrate these quaternion generators with string diagrams for the permutations. The reader can check that the permuations correspond to the permutation matrices constructed for the Klein Four Group. For example, the permutation for $I$ is $(12)(34)$ in cycle notation, the permutation for $J$ is $(13)(24)$ and the permutation for $K$ is $(14)(23).$ In the Figure
			we attach signs to each string of the permutation. These ``signed permutations'' act exactly as the products of vectors and permutations that we use for the iterants. One can see that the quaternions arise naturally from the Klein Four Group by attaching signs to the generating permutations as we have done in this Figure.

\begin{figure}
     \begin{center}
     \begin{tabular}{c}
     \includegraphics[width=6cm]{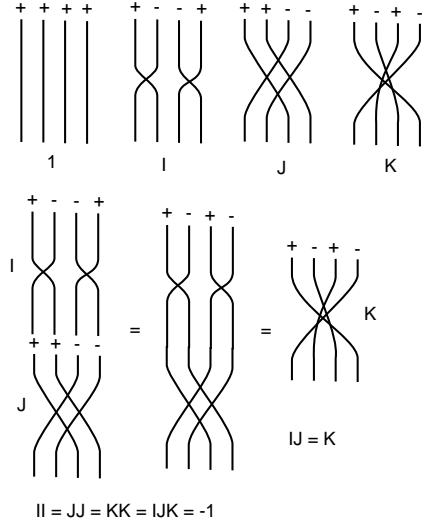}
     \end{tabular}
     \caption{\bf Quaternions From Klein Four Group}
     \label{fourgroup}
\end{center}
\end{figure}
\bigbreak

\item One can use the quaternions as a linear basis for $4 \times 4$ matrices just as our theorem would use the permutation matrices $1, A,B,C.$ If we restrict to real scalars $a,b,c,d$ such that $a^2 + b^2 + c^2 + c^2 = 1,$ then the set of matrices of the form $a1 + bI + cJ + dK$ is isomorphic to the group 
$SU(2).$
To see this, note that $SU(2)$ is the set of matrices with complex entries $z$ and $w$ with 
determinant $1$ so that $z \bar{z} + w \bar{w} = 1.$
$$M = \left(\begin{array}{cc}
			z& w\\
			 -\bar{w}&\bar{z}\\
			\end{array}\right).$$
Letting $z = a + bi$ and w = $c + di,$ we have
$$M =                 \left(\begin{array}{cc}
			a + bi& c + di\\
			 -c+di &a - bi\\
			\end{array}\right) = a
			\left(\begin{array}{cc}
			1& 0\\
			 0&1\\
			\end{array}\right) +b
			\left(\begin{array}{cc}
			i& 0\\
			 o&-i\\
			\end{array}\right) + c
			\left(\begin{array}{cc}
			0& 1\\
			 -1&0\\
			\end{array}\right)+d
			\left(\begin{array}{cc}
			0& i\\
			 i&0\\
			\end{array}\right).
			$$
If we regard $i = \sqrt{-1}$ as a commuting scalar, then we can write the generating matrices in terms
of size two iterants and obtain $$I=\sqrt{-1}\epsilon, J= \epsilon \eta, K= \sqrt{-1}\eta$$ as described in the previous section. IF we regard these matrices with complex entries as shorthand for $4 \times  4$ matrices with $i$ interpreted as a $2 \times 2$ matrix as we have done above, then these $4 \times 4$ matrices representing the quaternions are exactly the ones we have constructed in relation to the Klein Four Group.

Since complex numbers commute with one another, we could consider iterants whose values are in the complex numbers. This is just like considering matrices whose entries are complex numbers.
For this purpose we shall allow given  a version of $i$ that commutes with the iterant shift operator 
$\eta.$ Let this commuting $i$ be denoted
by $\iota.$ Then we are assuming that 

$$\iota^2 = -1$$ 
$$\eta \iota= \iota \eta$$
$$\eta^2 = +1.$$

We then consider iterant views of the form $[a + b\iota, c+ d\iota]$ and
$[a + b\iota , c + d\iota ]\eta = \eta[c + d\iota , a + b\iota ].$ In particular, we have
$\epsilon = [1,-1],$ and $i = \epsilon \eta$  is quite distinct from $\iota.$ Note, as before, that
$\epsilon \eta  = -\eta  \epsilon$ and that $\epsilon^2 = 1.$ Now let

$$I = \iota \epsilon$$ 
$$J = \epsilon \eta$$  
 $$K = \iota \eta.$$

We have used the commuting version of the square root of minus one in these definitions, and indeed we find the quaternions once more. 

$$I^2 =  \iota \epsilon \iota \epsilon =  \iota \iota \epsilon \epsilon = (-1)(+1) = -1,$$
$$J^2 =  \epsilon \eta \epsilon \eta =  \epsilon  (-\epsilon) \eta  \eta  = -1,$$
$$K^2 =   \iota \eta  \iota \eta=   \iota  \iota \eta \eta  = -1,$$
$$IJK = \iota \epsilon  \epsilon \eta \iota \eta  = \iota  1 \iota \eta \eta  = \iota \iota  = -1.$$

Thus
$$I^2 = J^2 = K^2 = IJK = -1.$$

This construction shows how the structure of the quaternions comes directly from the non-commutative structure of period two iterants. In other, words, quaternions can be represented by $2 \times  2$ matrices. This is the way it has been presented in standard language. The group $SU(2)$ of
$2 \times 2$  unitary matrices of determinant one is isomorphic to the quaternions of length one.

\item Similarly, $$H = [a,b] + [c + d\iota , c-d\iota ]\eta =
\left(\begin{array}{cc}
			a&c + d\iota\\
			  c-d\iota &b\\
			\end{array}\right).$$

represents a Hermitian $2 \times 2$ 
matrix and hence an observable for quantum processes mediated by
$SU(2).$ Hermitian matrices have real eigenvalues.
\bigbreak

If in the above Hermitian matrix form we take $a=T+X, b=T-X, c = Y, d=Z,$ then we obtain an iterant and/or matrix representation for a point in Minkowski spacetime. 
 $$H = [T+X,T-X] + [Y + Z\iota , Y-Z\iota ]\eta =
\left(\begin{array}{cc}
			T+X&Y + Z\iota\\
			  Y-Z\iota &T-X\\
			\end{array}\right).$$
Note that we have the formula $$Det(H) = T^2 - X^2 - Y^2 - Z^2.$$ It is not hard to see that the eigenvalues of $H$ are $T \pm \sqrt{X^2 + Y^2 + Z^2}.$ Thus, viewed as an observable, $H$ can observe the time and the invariant spatial distance from the origin of the event $(T,X,Y,Z).$ At least at this very elementary juncture, quantum mechanics and special relativity are reconciled.

\item Hamilton's Quaternions are generated by iterants, as discussed above, and we can express them
purely algebraicially by writing the corresponding permutations as shown below.

$$I = [+1,-1,-1,+1]s$$
$$J= [+1,+1,-1,-1]l$$
$$K= [+1,-1,+1,-1]t$$
where 

$$s =(12)(34)$$
$$l= (13)(24)$$
$$t =(14)(23).$$

Here we represent the permutations as products of transpositions $(ij).$ The transposition $(ij)$ interchanges $i$ and $j,$  leaving all other elements of $\{1,2,...,n \}$ fixed.

One can verify that 

$$I^2 = J^2 = K^2 = IJK = -1.$$
For example,

$$I^2 = [+1,-1,-1,+1]s [+1,-1,-1,+1]s$$
$$= [+1,-1,-1,+1][-1,+1,+1,-1]s s $$
$$= [-1,-1,-1,-1] $$
$$= -1.$$
and

$$IJ = [+1,-1,-1,+1]s [+1,+1,-1,-1]l$$
$$= [+1,-1,-1,+1][+1,+1,-1,-1] s l$$
$$= [+1,-1,+1,-1] (12)(34)(13)(24)$$
$$= [+1,-1,+1,-1] (14)(23)$$
$$= [+1,-1,+1,-1] t.$$

Nevertheless, we must note that making an iterant interpretation of
an entity like $I = [+1,-1,-1,+1]s$  is a conceptual departure from our original period two iterant (or cyclic period $n$) notion. Now we are considering iterants such as $[+1,-1,-1,+1]$ where the permutation group acts to produce other orderings of a given sequence. The iterant itself is not necessarily an oscillation. It can represent an implicate form that can be seen in any of its possible orders. These orders are subject to permutations that produce the possible views of the iterant. Algebraic structures such as the quaternions appear in the explication of such implicate forms.
\bigbreak

The reader will also note that we have moved into a different conceptual domain from an original emphasis in this paper on eigenform in relation to to recursion. That is, we take an {\it eigenform} to mean a fixed point for a transformation. Thus $i$ is an eigenform for $R(x) = -1/x.$
Indeed, each generating quaternion is an eigenform for the transformation $R(x) = -1/x.$
The richness of the quaternions arises from the closed algebra that arises with its infinity of eigenforms that satisfy the equation $U^2 = -1:$  $$U = aI + bJ + cK$$ where $a^2 + b^2 + c^2  = 1.$
This kind of significant extra structure in the eigenforms comes from paying attention to specific aspects of implicate and explicate structure, relationships with geometry and ideas and inputs from the perceptual, conceptual and physical worlds. Just as with our other examples  of phenomena arising in the course of the recursion, we see the same phenomena here in the evolution of matheamatical and theoretical physical structures in the course of the recursion that constitutes scientific conversation.

\item In all these examples, we have the opportunity to interpret the iterants as short hand for matrix algebra based on permutation matrices, or as indicators of discrete processes. The discrete processes become more complex in proportion to  the complexity of the groups used in the construction.
We began with processes of order two, then considered cyclic groups of arbitrary order, then the symmetric group $S_{3}$ in relation to $6 \times 6$ matrices, and the Klein Four Group in relation to the quaternions. In the case of the quaternions, we know that this structure is intimately related to rotations of three and four dimensional space and many other geometric themes. It is worth reflecting on the possible significance of the underlying discrete dynamics for this geometry, topology and related physics.

\end{enumerate}
\bigbreak

\section{The Iterant Algebra ${\cal A}_{n}$}
In this section, we will formulate relations with matrix algebra as follows.
Let $M$ be an $n \times n$ matrix over a ring $F.$ Let $M=(m_{ij})$ denote the 
matrix entries. Let $\pi$ be an element of the symmetric group $S_{n}$ so that 
$\pi_1, \pi_2, \cdots , \pi_n$ is a permuation of $1,2,\cdots, n.$ Let 
$v = [v_1, v_2, \cdots, v_n]$ denote a vector with these components.
Let $\Delta(v)$ denote the diagonal matrix whose $i-th$ diagonal entry is $v_i.$
Let $v^{\pi} = [v_{\pi_1},\cdots, v_{\pi_n}].$ Let $\Delta^{\pi}(v) = \Delta(v^{\pi}).$
Let $\Delta$ denote any diagonal matrix and $\Delta^{\pi}$ denote the corresponding permuted
diagonal matrix as just described. Let $P[\pi]$ denote the permutation
matrix obtained by taking the $i-th$ row of $P[\pi]$ to be the
$\pi_{i} -th$ row of the identity matrix. Note that $P[\pi]\Delta = \Delta^{\pi}P[\pi].$
For each element $\pi$ of $S_{n}$ define the vector $v(M,\pi) = [m_{1\pi_1},\cdots,m_{n\pi_n}]$
and the diagonal matrix $\Delta[M]_{\pi} = \Delta(v(M, \pi)).$
\bigbreak

Given an $n \times n$ permutation matrix $P[\sigma]$ and a diagonal matrix $D,$ the matrix
$DP[\sigma]$ has the entries of $D$ in those places where there were $1$'s in $P[\sigma].$
Let $a(D) = [D_{11},D_{22},\cdots, D_{nn}]$ be the {\it iterant associated with $D.$}
\bigbreak

Consider $n$-tuples $a = [a_{1},\cdots, a_{n}]$ where $a_{i} \in F,$ and let the symmetric group $S_{n}$ act on these $n$-tuples by permutation of the coordinates. Let
$e_{i}$ denote such an $a$ where $a_{i} = 1$ and all the other coordinates are zero.
Let $a^{\sigma} = [a_{\sigma (1)}, \cdots, a_{\sigma (n)}]$ be the vector obtained by letting $\sigma \in S_{n}$ act on $a.$ Note that $$a = \sum_{k=1}^{k=n} a_{k}e_{k}.$$
Define the
{\it iterant algebra} ${\cal A}_{n}$ to be the module over $F$ with basis
${\cal B} = \{ e_{i} \gamma | i = 1, \cdots n; \gamma \in S_{n}\}$ where the algebra structure is given
by $$(a \sigma)(b \tau) = ab^{\tau} (\sigma \tau).$$ We see that  
$$dim({\cal A}_{n}) = n \times n! = n^2 \times (n-1)!.$$ 
\bigbreak

Let ${Matr}_{n}$ denote the set of $n \times n$ matrices over the ring $F.$ 
Note that since the permutation representation used for $S_{n}$ is the same as the right regular representation only for $n=2,$ we have that
${\cal A}_{2} \simeq {Matr}_{2} \simeq \mathbb{V}ect_{2}(S_{2}, \mathbb{F}) ,$ as defined in the previous section.
For other values of $n$ we will analyze the relationships of these rings. 
\bigbreak

 Let  $$p: {\cal A}_{n} \longrightarrow Matr_{n}$$ via $$p (a \sigma) = \Delta(a) P[\sigma]$$
 where $\Delta(a)$ is the diagonal matrix associated with the iterant $a$ and $P[\sigma]$ is the 
 permutation matrix associated with the permuation $\sigma.$ Then $\rho$ is a matrix representation of the iterant algebra $ {\cal A}_{n}.$ This is not a faithful representation. Note that if
 $\sigma(i) = \tau(i)$ for permuations $\sigma$ and $\tau,$ then $\rho(e_{i} \sigma) = \rho(e_{i} \tau).$
 It remains to be seen how to form the full representation theory for the algebra ${\cal A}_{n}.$
 This will be a generalization of the representation theory for the group algebra of the symmetric group, which is ${\cal A}_{1}.$
 \bigbreak

A reason for discussing these formulations of matrix algebra in the present context is that one sees that 
matrix algebra is generated by the simple operations of juxtaposed addition and multiplication, and by
the use of permutations as operators. These are unavoidable discrete elements, and so the operations of matrix
algebra can be motivated on the basis of discrete physical ideas and non-commutativity. The richness of  
continuum formulations, infinite matrix algebra, and symmetry grows naturally out of 
finite matrix algebra and hence out of the discrete.
\bigbreak

\noindent {\bf Theorem.} Let $M$ denote an $n \times n$ matrix with entries in a ring (associative not necessarily commutative)  with unit. Then 
$$M = \frac{1}{(n-1)!}\Sigma_{\pi \in S_n} \Delta[M]_{\pi} P[\pi].$$
This means that ${\cal M}_{n}$ can be embedded in ${\cal A}_{n}$, for we have 
the map $i:{\cal M}_{n} \longrightarrow  {\cal A}_{n}$ defined by
$$i(M) =  \frac{1}{(n-1)!}\Sigma_{\pi \in S_n} v(M, \pi)\pi$$ and $$p \circ i = 1_{Matr_{n}}.$$
This implies that $${\cal A}_{n} \simeq {\cal K}_{n} \oplus Matr_{n}$$ where ${\cal K}_{n}$ is the kernel of $p.$
\bigbreak

\noindent {\bf Proof.} 
Let $\delta_{ij}$ denote the Kronecker delta, equal to $1$ when $i=j$ and equal to $0$ otherwise.
The matrix product $\Delta[M]_{\pi} [\pi]$ is given as follows.
\begin {enumerate}
\item $(\Delta[M]_{\pi} [\pi])_{ij} = A_{i \pi_{i}} =A_{ij} \delta_{j\pi_{i}}$ if $j = \pi_{i}$.
\item $(\Delta[M]_{\pi} [\pi])_{ij} = 0$ if $j \ne \pi_{i}$.
\end{enumerate}

\noindent This follows from the fact that 
$$\Delta[M]_{\pi} = \left(\begin{array}{cccc}
			A_{1\pi_{1}}&0& \cdots &0\\
			0&A_{2\pi_{2}}& \cdots &0\\
			~&~& \cdots &~\\
			0& \cdots &0 &A_{n\pi_{n}}\\
			\end{array}\right).$$

\noindent We abbreviate $$ \Delta[M]_{\pi} =  \Delta_{\pi}.$$
Hence,
$$(\sum_{\pi \in S_{n}} \Delta_{\pi} [\pi]))_{ij} = \sum_{\pi \in S_{n}}(\Delta_{\pi} [\pi])_{ij} $$
$$= \sum_{\pi \in S_{n}} A_{ij} \delta_{j\pi_{i}} = A_{ij} \sum_{\pi \in S_{n}} \delta_{j \pi_{i}}.$$
$\sum_{\pi \in S_{n}} \delta_{j \pi_{i}} = $ $($ the number of permutations of $123 \cdots n$ with 
$\pi_{i} = j) = (n-1)!.$
This  completes the proof of the Theorem. //
\bigbreak

\noindent Note that the theorem expresses any square 
matrix as a sum of products of diagonal matrices and permutation matrices. Diagonal matrices add and
multiply by adding and multiplying their corresponding entries. They are acted upon by permutations as 
described above.  This is a full generalization of the case $n=2$ described in the last section.
\bigbreak

For example, we have the following expansion of a $3 \times 3$ matrix:
$$\left(\begin{array}{ccc}
a&b&c\\
d&e&f\\
g&h&k
\end{array}\right) = \frac{1}{2!}[
\left(\begin{array}{ccc}
a&0&0\\
0&e&0\\
0&0&k
\end{array}\right) +
\left(\begin{array}{ccc}
0&b&0\\
0&0&f\\
g&0&0
\end{array}\right) +
\left(\begin{array}{ccc}
0&0&c\\
d&0&0\\
0&h&0
\end{array}\right) +$$
$$\left(\begin{array}{ccc}
0&0&c\\
0&e&0\\
g&0&0
\end{array}\right)+
\left(\begin{array}{ccc}
0&b&0\\
d&0&0\\
0&0&k
\end{array}\right)+
\left(\begin{array}{ccc}
a&0&0\\
0&0&f\\
0&h&0
\end{array}\right)].
$$
Here, each term factors as a diagonal matrix multiplied by a permutation matrix as in
$$\left(\begin{array}{ccc}
a&0&0\\
0&0&f\\
0&h&0
\end{array}\right) = \left(\begin{array}{ccc}
a&0&0\\
0&f&0\\
0&0&h
\end{array}\right) \left(\begin{array}{ccc}
1&0&0\\
0&0&1\\
0&1&0
\end{array}\right).$$
It is amusing to note that this theorem tells us that up to the factor of $1/(n-1)!$ a unitary matrix
that has unit complex numbers as its entries is a sum of simpler unitary transformations factored into
diagonal and permutation matrices. In quantum computing parlance, such a unitary matrix is a sum of products of
phase gates and products of swap gates (since each permutation is a product of transpositions).
\bigbreak 

Abbreviating a diagonal matrix by the ``iterant`` $\Delta [a,b,c]$, we write
$$\left(\begin{array}{ccc}
a&0&0\\
0&b&0\\
0&0&c
\end{array}\right) = \Delta[a,b,c].$$ Then we can write the entire decomposition of the $3 \times 3$ matrix in the form shown below.
$$(2!) \left(\begin{array}{ccc}
a&b&c\\
d&e&f\\
g&h&k
\end{array}\right) = 
\Delta[a,e,k] \left(\begin{array}{ccc}
1&0&0\\
0&1&0\\
0&0&1
\end{array}\right) +
\Delta[b,f,g] \left(\begin{array}{ccc}
0&1&0\\
0&0&1\\
1&0&0
\end{array}\right) +
\Delta[c,d,h] \left(\begin{array}{ccc}
0&0&1\\
1&0&0\\
0&1&0
\end{array}\right) +$$
$$
\Delta[a,f,h] \left(\begin{array}{ccc}
1&0&0\\
0&0&1\\
0&1&0
\end{array}\right)+
\Delta[c,e,g]  \left(\begin{array}{ccc}
0&0&1\\
0&1&0\\
1&0&0
\end{array}\right)+
\Delta[b,d,k] \left(\begin{array}{ccc}
0&1&0\\
1&0&0\\
0&0&1
\end{array}\right).
$$

Thus
$$(2!) \left(\begin{array}{ccc}
a&b&c\\
d&e&f\\
g&h&k
\end{array}\right) = 
\Delta[a,e,k]  + \Delta[b,f,g]  \rho  + \Delta[c,d,h] \rho^{2} +\Delta [a,f,h] \tau +\Delta [c,e,g]\rho \tau  +\Delta [b,d,k] \rho^{2} \tau$$
$$=\Delta [a,e,k]  + \Delta[b,f,g]  \rho  +\Delta [c,d,h] \rho^{2} +\Delta [a,f,h] \tau_{1}+\Delta [c,e,g] \tau_{2}  + \Delta[b,d,k] \tau_{3}. $$
Here $\rho = (123)$ and $\tau = \tau_{1}=(23), \tau_{2}= (13), \tau_{3} = (12)$ in the standard cycle notation for permutations. {\it We write abstract permutations and the corresponding permutation matrices interchangeably.}
The reader can easily spot the matrix definitions of these generators of $S_{3}$ by comparing the last equation to previous equation.
\bigbreak

Note that in terms of the mapping $p: {\cal A}_{3} \longrightarrow Matr_{3},$ we have that 
$$p([a,e,k]  + [b,f,g]  \rho  + [c,d,h] \rho^{2} + [a,f,h] \tau_{1}+  [c,e,g] \tau_{2}  +  [b,d,k] \tau_{3}) = (2!) \left(\begin{array}{ccc}
a&b&c\\
d&e&f\\
g&h&k
\end{array}\right).$$
\bigbreak

In this form, matrix multiplication disappears and we can calculate sums and products entirely with iterants and the action of the permutations on these iterants. The reader will note immediately that the 
full algebra ${\cal A}_{3}$ for iterants of size $[a,b,c]$ is larger and more general than $3 \times 3$ matrix algebra. We let the entries in the iterants belong to a field $F.$
The most general element in this algebra is given by the formula
$${\cal I} = [a,b,c]  + [d,e,f]  \rho  + [g,h,i] \rho^{2} + [j,k,i] \tau_{1}  + [m,n,o] \tau_{2} + [p,q,r] \tau_{3}.$$
where $a,b,\cdots r$ are elements of $F.$ We do not assume that the group elements are 
represented by matrices, but we do have them act on the iterants $[x,y,z]$ by permuting the coordinates.
Letting $e_{1}=[1,0,0], e_{2}=[0,1,0], e_{3}=[0,0,1],$ we have that $\{ e_{i} g | i = 1,2,3; g \in S_{3}\}$ is a basis for ${\cal A}_{3}$ over the field $F.$
Thus the dimension of this algebra
is $3 \times 3! = 18.$
\bigbreak

We have the exact sequence $$0 \longrightarrow {\cal K}_{n} \longrightarrow {\cal A}_{n} \longrightarrow Matr_{n} \longrightarrow 0,$$ with $p: {\cal A}_{n} \longrightarrow Matr_{n}$ and 
$i: Matr_{n}  \longrightarrow   {\cal A}_{n}.$ Here are some examples of elements of the kernel 
${\cal K}_{n}$ of $p.$ Let $x = [1,0,0] - [1,0,0](23) \in {\cal A}_{3}.$ Then it is easy to see that $p(x)=0.$
$x$ itself is a non-trivial element of ${\cal A}_{3},$ Note that $x^2 = 2x,$ so $x$ is not nilpotent.
We know from the fundamental classification theorem for associative algebras \cite{Littlewood} that 
${\cal A}_{n}/N$ (where N is the subalgebra of properly nilpotent elements of ${\cal A}_{n}$) is isomorphic to a full matrix algebra. Thus we see that the decomposition that we have given for 
${\cal A}_{n}$ is distinct from the one obtained by removing the nilpotent elements. It remains to classify the nilpotent subalgebra of  ${\cal A}_{n}.$ We shall return to this question in a sequel to this paper. 
\bigbreak

Here is a final example of an element in the kernel of $p.$ Consider the matrix
$$M =  \left(\begin{array}{ccc}
a&b&c\\
c&a &b\\
b&c&a
\end{array}\right).$$ We can write this matrix quite simply as a sum of scalars times three permutation matrices generating the cyclic group of order three.
$$M =  a\left(\begin{array}{ccc}
1&0&0\\
0&1&0\\
0&0&1
\end{array}\right)+
b\left(\begin{array}{ccc}
0&1&0\\
0&0&1\\
1&0&0
\end{array}\right)+
c\left(\begin{array}{ccc}
0&0&1\\
1&0&0\\
0&1&0
\end{array}\right).
$$
However, our mapping $i:Matr_{3} \longrightarrow {\cal A}_{3}$ includes terms for all the permutation matrices and adds, essentially, three more terms to this formula. 
$$2 \times i(M) = a1 + b(123) + c(132) + [c,a,b](13) + [b,c,a](12) + [a,b,c](23).$$
Consequently, $$y = a1 + b(123) + c(132) -[c,a,b](13) - [b,c,a](12) - [a,b,c](23)$$
belongs to the kernel of the mapping $p.$
\bigbreak
 
 \noindent {\bf Lemma.} The kernel ${\cal K}_{3}$ of the mapping $p:{\cal A}_{3} \longrightarrow Matr_{3}$ consists in the elements 
 $$[x,y,z]+ [-x,w,t] \tau_{1} + [r,-y,s] \tau_{2} + [p,q,-z] \tau_{3} + [-p,-w,-s] \rho + [-r,-q,-t] \rho^{2}.$$
 \smallbreak
 
 \noindent {\bf Proof.} We leave this proof to the reader.//
\bigbreak

 \noindent {\bf Proposition.} The kernel ${\cal K}_{n}$ of the mapping $p:{\cal A}_{n} \longrightarrow Matr_{n}$ consists in the elements $$\alpha = \Sigma_{\sigma \in S_{n}} a_{\sigma} \sigma$$ such that for all $i,j$ with  $1 \le i, j \le n,$ $$\Sigma_{\sigma: \sigma(i) =j} (a_{\sigma})_{i} = 0.$$ Thus we have that   ${\cal A}_{n}/{\cal K}_{n}$ is isomorphic to the full matrix algebra $Matr_{n}.$
 \smallbreak

\noindent{\bf Proof.} The proposition follows from the fact that $p(\alpha) = A$ where 
$$A_{i,j}= \Sigma_{\sigma: \sigma(i) =j} (a_{\sigma})_{i}.$$ //
\bigbreak

In a subsequent paper we shall turn to the apparently more  difficult problem of fully understanding the 
structure of the algebras ${\cal A}_{n}$ for $n \ge 3.$ Here we have seen that the fact that the kernel of the mapping $p$ is non-trivial means that there is often a choice in making an iterant representation for a given matrix or for an algebra of matrices. In many applications, certain underlying permutation matrices stand out and so suggest themselves as a basis for an iterant representation. This is the case for the quaternions, as we have seen. It is also the case for the Dirac matrices and other matrices that occur in physical applications. We shall discuss some of these examples below.
\bigbreak

 \bigbreak
 
 \section{The Square Root of Minus One is a Clock}
The purpose of this section is to place $i,$ the  square root of minus one, and its algebra in a context of discrete recursive systems. We begin by starting with a simple periodic process that is associated directly with the classical attempt to solve for $i$ as a solution to a quadratic equation. We take the point
of view that solving $x^2 = ax + b$ is the same (when $x \ne 0$) as solving 
$$ x = a + b/x, $$ and hence is a matter of finding a fixed point. In the case of $i$ we have
$$ x^{2} = -1 $$ and so desire a fixed point $$ x = -1/x. $$ There are no real numbers that are fixed points for this operator and so we consider the oscillatory process generated by
$$R(x) = -1/x. $$
The fixed point would satisfy
$$i = -1/i$$
and multiplying, we get that 
$$ii = -1.$$
On the other hand the iteration of R yields
$$ 1, R(1) = -1 , R(R(1)) = +1 , R(R(R(1))) = -1, +1,-1,+1,-1, \cdots .$$
The square root of minus one is a perfect example of an eigenform that occurs in a new and wider domain than the original context in which its recursive process arose. The process has no fixed point in the original domain.
\smallbreak

Looking at the oscillation between $+1$ and $-1,$ we see that there are naturally two phase-shifted viewpoints.  We denote these two views of the  oscillation by $[+1,-1]$ and$ [-1,+1].$ 
These viewpoints correspond to whether one regards the oscillation at time zero as starting with $+1$ or with $ -1.$ See Figure 1.

\begin{figure}
     \begin{center}
     \begin{tabular}{c}
     \includegraphics[width=6cm]{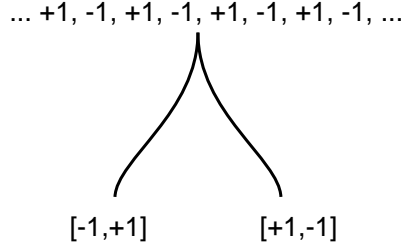}
     \end{tabular}
     \caption{\bf A Basic Oscillation}
     \label{Figure 1}
\end{center}
\end{figure}

We shall let $I\{+1,-1\}$ stand for an undisclosed alternation or ambiguity between $+1$ and $-1$  and call $ I\{+1,-1\}$  an iterant. There are two iterant views: $[+1,-1]$ and $[-1,+1]. $
\smallbreak

Given an iterant $[a,b],$ we can think of $[b,a]$ as the same process with a shift of one time step.
These two iterant views, seen as points of view of an alternating process, will become the square roots of negative unity, $i$ and $-i.$
\smallbreak

We introduce a temporal shift operator $ \eta$  such that 
$$[a,b]\eta  = \eta  [b,a]$$ and $$\eta \eta  = 1$$ 
for any iterant $[a,b],$  so that concatenated observations can include a time step of one-half period of the process 
$$\cdots abababab \cdots.$$ 
We combine iterant views term-by-term as in 
$$[a,b][c,d] = [ac,bd]. $$
We now define i by the equation
$$i = [-1,1] \eta . $$
This makes $i $ both a value and an operator that takes into account a step in time.
\smallbreak

We calculate
$$ii = [-1,1] \eta [-1,1] \eta  = [-1,1][1,-1]\eta \eta  = [-1,-1] = -1.$$
Thus we have constructed the square root of minus one by using an iterant viewpoint.  In this view $i$ represents a discrete oscillating temporal process and it is an eigenform for $R(x) = -1/x,$ participating in the algebraic structure of the complex numbers. In fact the corresponding algebra structure of 
linear combinations $[a,b] + [c,d]\eta$ is isomorphic with $2 \times 2$ matrix algebra and iterants can be used to construct $n \times n$ matrix algebra, as we have already discussed in this paper.
\smallbreak

\noindent {\bf The Temporal Nexus.}
{\em We take as a matter of principle that the usual real variable $t$ for time is better represented as $it$ so that time is seen to be a process, an observation and a magnitude all at  once. }
This principle of ``imaginary time" is justified by the eigenform approach to the structure of time and the structure of the square root of minus one.
\smallbreak

As an example of the use of the Temporal Nexus, consider the 
expression $x^2 + y^2 + z^2 + t^2,$ the square of the Euclidean distance of a point $(x,y,z,t)$ from the origin in Euclidean four-dimensional space. Now replace $t $ by $it,$ and find 
$$x^2 + y^2 + z^2 + (it)^2 = x^2 + y^2 + z^2 - t^2,$$
the squared distance in hyperbolic metric for special relativity. By replacing t by its process operator value $it$  we make the transition to the physical mathematics of special relativity.
\smallbreak

  In this section we shall first apply this 
idea to Lorentz
 transformations, and then generalize it to other contexts.  
 \vspace{3mm}
 
 So, to work: We have $$[t-x,t+x] = [t,t] + [-x,x] = t[1,1] + x[-1,1].$$  
 Since $[1,1][a,b] = [1a,1b] = [a,b]$  and $[0,0][a,b]= [0,0]$, we shall 
write
 $$1=[1,1]$$ and $$0=[0,0].$$ Let  $$\sigma = [-1,1].$$  $\sigma$ is a 
significant 
 iterant that we shall refer to as a {\em polarity}.  Note that $$\sigma 
\sigma = 1.$$
 Note also that $$[t-x,t+x] = t + x\sigma.$$ Thus the points of 
spacetime form an
 algebra analogous to the complex numbers whose elements are of 
the form $t+x\sigma$
 with $\sigma \sigma = 1$  so that 
 $$(t+x\sigma)(t'+x'\sigma) = tt'+xx' +(tx'+xt')\sigma.$$
 In the case of the Lorentz transformation it is easy to see the 
elements of the form
 $[k,k^{-1}]$ translate into elements of the form

 $$T(v) = [(1+v)/\sqrt{(1-v^{2})}, (1-v)/\sqrt{(1-v^{2})}] = [k,k^{-1}].$$
 
 Further analysis shows that $v$ is the relative velocity of the two 
reference frames in
 the physical context. Multiplication now yields the usual form of the 
Lorentz transform
 $$T_{k}(t + x\sigma) = T(v)(t+x\sigma)$$ 
 $$= (1/\sqrt{(1-v^{2})} -v\sigma/\sqrt{(1-v^{2})})(t+x\sigma)$$ 
 $$=(t-xv)/\sqrt{(1-v^{2})} + (x-vt)\sigma/\sqrt{(1-v^{2})}$$ 
 $$= t' + x'\sigma.$$
 \vspace{3mm}

 The algebra that underlies this iterant presentation of special 
relativity is a relative  
 of the complex numbers with a special element $\sigma$ of square 
one rather than minus
 one ($i^{2} = -1$  in the complex numbers).  
 \vspace{3mm}

 
\section{The Wave Function in Quantum Mechanics and The Square Root of Minus One}
One can regard  a wave function such as $\psi (x,t) =exp(i(kx - wt))$ as containing a micro-oscillatory system with the  special synchronizations of the iterant view $ i = [+1,-1]\eta$ . It is these synchronizations that make the big eigenform of the exponential work correctly with respect to differentiation, allowing it to create the appearance of rotational behaviour, wave behaviour and the semblance of the continuum. In other words, we are suggesting that one can take a temporal view of the well-known equation of Euler: $$e^{i \theta} = cos(\theta) + i sin(\theta)$$ by regarding the $i$
in this equation as an iterant, as a discrete oscillation between $-1$ and $+1.$ 
One can blend the classical geometrical view of the complex numbers with the iterant view by thinking of a point that orbits the origin of the complex plane, intersecting the real axis periodically and producing, in the real axis, a periodic oscillation in relation to its orbital movement in the two dimensional space.
The special synchronization is the algebra of the time shift embodied in 
$$\eta \eta = 1$$ and 
$$[a,b] \eta = \eta [b,a]$$
that makes the algebra of $i = [1,-1]\eta$ imply that $i^2 = -1.$
This interpretation does not change the formalism of these complex-valued functions, but it does change
one's point of view and we now show how the properties of $i$ as a discrete
dynamical systerm are found in any such system.
\smallbreak

\subsection{Time Series and Discrete Physics}
We have just reformulated the complex numbers and expanded the context of matrix algebra to an interpretation of $i$ as an oscillatory process and matrix elements as combined spatial and temporal oscillatory processes (in the sense that $[a,b]$ is not affected in its order by a time step, while 
$[a,b] \eta$ includes the time dynamic in its interactive capability, and $2 \times  2$ matrix algebra is the algebra of iterant views $[a,b] + [c,d] \eta $). 
\smallbreak

We now consider elementary discrete physics in one dimension. Consider a time series of positions
$$x(t):  t = 0, \Delta t, 2\Delta t, 3\Delta t, \cdots . $$
We can define the velocity $v(t)$ by the formula 
$$v(t) = (x(t+ \Delta t) - x(t))/\Delta t = Dx(t)$$ 
where $D$ denotes this discrete derivative. In order to obtain $v(t)$ we need at least one tick  $\Delta t$ of the discrete clock.  Just as in the iterant algebra, we need a time-shift operator to handle the fact that once we have observed $v(t),$ the time has moved up by one tick. 
\bigbreak

\noindent{\bf We adjust the discrete derivative.}
We shall add an operator J that in this context accomplishes the time shift:
$$x(t)J = Jx(t+\Delta t).$$
We then redefine the derivative to include this shift:
$$Dx(t) = J(x(t+ \Delta t) - x(t))/\Delta t .$$
This readjustment of the derivative rewrites it so that the temporal properties of successive
observations are handled automatically.
\smallbreak

\noindent{\bf Discrete observations do not commute.}
Let $A$ and $B$ denote quantities that we wish to observe in the discrete system.
Let $AB$ denote the result of first observing $B$ and then observing $A.$
The result of this definition is that a successive observation 
of the form $x(Dx)$ is distinct from an observation of the form
$(Dx)x.$ In the first case, we first observe the velocity at time $t$, and then $x$ is measured at $t + \Delta t $. In the second case, we measure $x$ at $t $ and then 
measure the velocity.
\smallbreak

We measure the difference between these two results by taking a commutator $$[A,B] = AB - BA$$ and we get the following computations where we write  $\Delta x = x(t+ \Delta t) - x(t).$
$$x(Dx) = x(t)J( x(t+ \Delta t) - x(t)) = Jx(t+ \Delta t)( x(t+ \Delta t) - x(t)).$$
$$(Dx)x = J(x(t+ \Delta t) - x(t))x(t).$$
$$[x,Dx] = x(Dx) - (Dx)x  = (J/\Delta t)(x(t+ \Delta t) - x(t))^2  = J (\Delta x)^2/\Delta t$$
This final result is worth recording:
$$[x,Dx] = J ( \Delta x)^2/\Delta t.$$
From this result we see that the commutator of $x$ and$ Dx$ will be constant if $(\Delta x)^2/\Delta t = K$ is a constant. For a given time-step,  this
means that $$(\Delta x)^2 = K \Delta t$$  so that  $$\Delta x =  \pm \sqrt{(K \Delta t )}$$This is a Brownian process with diffusion constant equal to $K.$
\smallbreak

Thus we arrive at the result that any discrete process viewed in this framework of discrete observation has the basic commutator $$[x,Dx] = J ( \Delta x)^2/\Delta t, $$ generalizing a Brownian process and 
containing the factor $ ( \Delta x)^2/\Delta t $ that corresponds to the classical diffusion constant.
It is worth noting that the adjusment that we have made to the discrete derivative makes it into a commutator as follows:
$$Dx(t) = J(x(t+ \Delta t) - x(t))/\Delta t  = (x(t)J - Jx(t))\Delta t = [x(t), J]/\Delta t.$$
By replacing discrete derivatives by commutators we can express discrete physics in many variables in a context of non-commutative algebra. See \cite{KN:QEM,KN:Dirac,NonCom,ST,Aspects,Boundaries,NCW,Fauser,Noyes} for more on this point of view.
\smallbreak

We now use the temporal nexus (the square root of minus one as a clock) and 
rewrite these commutators to match quantum mechanics.
\smallbreak

\subsection{Simplicity and the Heisenberg Commutator}

Finally, we arrive at the simplest place. 
Time and the square root of minus one  are inseparable in the temporal nexus. 
The square root of minus one is a symbol and algebraic operator for the simplest oscillatory process.
As a symbolic form, i is an eigenform satisfying the equation
$$i = -1/i.$$
One does not have an increment of time all alone as in classical $t. $
One has $it,$ a combination of an interval and the elemental dynamic that is time. With this understanding, we can return to the commutator for a discrete process and use $it$ for the temporal increment.
\smallbreak

We found that discrete observation led to the commutator equation
$$[x,Dx] = J (\Delta x)^2/ \Delta t$$
which we will simplify to
$$[q, p/m] = (\Delta x)^2/\Delta t.$$
taking $q$ for the position $x$ and $p/m$  for velocity, the time derivative of position and ignoring the
time shifting operator on the right hand side of the equation.
\smallbreak

Understanding that $\Delta t$ should be replaced by$ i \Delta t,$  and that, by comparison with the physics of a process at the Planck scale one can take
$$ (\Delta x)^2/\Delta t = \hbar /m, $$
we have
$$[q, p/m] = (\Delta x)^2/i \Delta t = -i \hbar /m,$$
whence
$$[p,q] = i\hbar,$$
and we have arrived at Heisenberg's fundamental relatiionship between position and momentum. This mode of arrival is predicated on the recognition that only  $it$ represents a true interval of time.
In the notion of time there is an inherent clock or an inherent shift of phase that is making a synchrony in our ability to observe, a precise dynamic beneath the apparent dynamic of the observed process. Once this substitution is made, once the correct imaginary value is placed in the temporal circuit, the patterns of quantum mechanics appear. In this way, quantum mechanics can be seen to emerge from the discrete.
\smallbreak

The problem that we have examined in this section  is the problem to understand the nature of  quantum mechanics. In fact, we hope that the problem is seen to disappear the more we enter into the present viewpoint. A viewpoint is only on the periphery. The iterant from which the viewpoint emerges is in a superposition of indistinguishables, and can only be approached by varying the 
viewpoint until one is released from the particularities that a point of view contains.
\bigbreak

\section{Clifford Algebra, Majorana Fermions and Braiding}
Recall fermion algebra. One has fermion annihiliation operators $\psi$ and their
conjugate creation operators $\psi^{\dagger}.$
One has $\psi^{2} = 0 = (\psi^{\dagger})^{2.}$
There is a fundamental commutation relation
$$\psi \psi^{\dagger} + \psi^{\dagger} \psi = 1.$$
If you have more than one of them say $\psi$ and $\phi$,
then they anti-commute:
$$\psi \phi = - \phi \psi.$$
The Majorana fermions $c$ that satisfy $c^{\dagger} = c$ so that they
are their own anti-particles. There is a lot of interest in these as
quasi-particles and they are related to braiding and to topological
quantum computing.  A group of researchers  \cite{Kouwenhouven}
 claims, at this writing, to
have found quasiparticle Majorana fermions in edge effects in nano-wires.
(A line of fermions could have a Majorana fermion happen non-locally from
one end of the line to the other.) The Fibonacci model that we discuss is also based on
Majorana particles, possibly related to collecctive electronic excitations. 
If $P$ is a Majorana fermion particle, then $P$ can interact with itself to either produce itself or to annihilate itself. This is the simple ``fusion algebra" for this particle. One can write
$P^2 = P + 1$ to denote the two possible self-interactions the particle $P.$
The patterns of interaction and braiding of such a particle $P$ give
rise to the Fibonacci model.\
\bigbreak

\begin{figure}
     \begin{center}
     \begin{tabular}{c}
     \includegraphics[height=7cm]{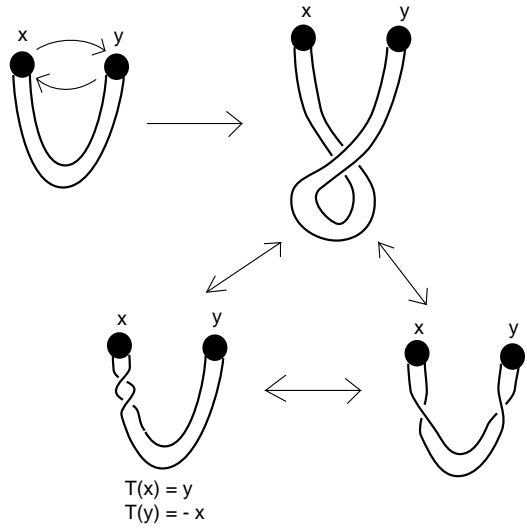}
     \end{tabular}
     \end{center}
     \caption{\bf Braiding Action on a Pair of Fermions}
     \label{ex}
     \end{figure}

Majoranas are related to standard fermions as follows:
The algebra for Majoranas is $c = c^{\dagger}$ and $cc' = -c'c$ if $c$ and $c'$ are
distinct Majorana fermions with  $c^{2}= 1$ and  $c'^{2}= 1.$
One can make a standard fermion from two Majoranas via
$$\psi = (c + ic')/2,$$
$$\psi^{\dagger} = (c -ic')/2.$$
Similarly one can
mathematically make two Majoranas from any single fermion.
Now if you take a set of Majoranas
$$\{ c_1, c_2, c_3, \cdots , c_n \}$$
then there are natural braiding operators that act on the vector space with
these $c_k$ as the basis. The operators are mediated by algebra elements
$$\tau_{k} =(1 + c_{k+1} c_{k})/\sqrt{2},$$
$$\tau_{k}^{-1} = (1 - c_{k+1} c_{k})/\sqrt{2}.$$
Then the braiding operators are
$$T_{k}: Span \{c_1,c_2,\cdots, ,c_n \} \longrightarrow Span \{c_1,c_2,\cdots, ,c_n \}$$
via 
$$T_{k}(x) = \tau_{k} x \tau_{k}^{-1}.$$
The braiding is simply:
$$T_{k}(c_{k}) = c_{k+1},$$
$$T_{k}(c_{k+1}) = - c_{k},$$
and $T_{k}$ is the identity otherwise.
This gives a very nice unitary representaton of the Artin braid group and
it deserves better understanding. See Figure~\ref{ex} for an illustration of this braiding of Fermions in relation to the topology of a belt that connects them. The relationship with the belt is tied up with the fact that in quantum mechanics we must represent rotations of three dimensional space as unitary transformations. See \cite{Kauff:KP} for more about this topological view of the physics of Fermions. In the Figure, we see that the belt does not know which of the two Fermions to annoint with the phase change, but the clever algebra above makes this decision. There is more to be done in this domain.
\bigbreak

     \bigbreak

It is worth noting that a triple of Majorana fermions say $a,b,c$ gives rise
to a representation of the quaternion group. This is a generalization of
the well-known association of Pauli matrices and quaternions.
We have $a^2 = b^2 = c^2 = 1$ and they anticommute.
Let $I = ba, J = cb, K = ac.$
Then $$I^2 = J^2 = K^2 = IJK = -1,$$ giving the quaternions.
The operators
$$A = (1/\sqrt{2})(1 + I)$$
$$B = (1/\sqrt{2})(1 + J)$$
$$C = (1/\sqrt{2})(1 + K)$$
braid one another: $$ABA = BAB,BCB =CBC, ACA = CAC.$$
This is a special case of the braid group representation described above
for an arbitrary list of Majorana fermions.
These braiding operators are entangling and so can be used for universal
quantum computation, but they give only partial topological quantum
computation due to the interaction with single qubit operators not
generated by them.
\bigbreak

Recall that in discussing the beginning of iterants, we introduce a {\it temporal shift operator} $ \eta$  such that 
$$[a,b]\eta  = \eta  [b,a]$$ and $$\eta \eta  = 1$$ 
for any iterant $[a,b],$  so that concatenated observations can include a time step of one-half period of the process 
$$\cdots abababab \cdots.$$ 
We combine iterant views term-by-term as in 
$$[a,b][c,d] = [ac,bd]. $$
We now define i by the equation
$$i = [1,-1] \eta . $$
This makes $i $ both a value and an operator that takes into account a step in time.
\smallbreak

We calculate
$$ii = [1,-1] \eta [1,-1] \eta  = [1,-1][-1,1]\eta \eta  = [-1,-1] = -1.$$
Thus we have constructed a square root of minus one by using an iterant viewpoint.  In this view $i$ represents a discrete oscillating temporal process and it is an eigenform for $T(x) = -1/x,$ participating in the algebraic structure of the complex numbers. In fact the corresponding algebra structure of 
linear combinations $[a,b] + [c,d]\eta$ is isomorphic with $2 \times 2$ matrix algebra and iterants can be used to construct $n \times n$ matrix algebra, as we have already discussed.
\smallbreak

Now we can make contact with the algebra of the Majorana fermions. Let $e = [1,-1].$ Then we have
$e^2 = [1,1] = 1$ and $e \eta  =  [1,-1] \eta  =  [-1,1] \eta  = - e \eta.$
Thus we have $$e^2 =1,$$ $$\eta^2 = 1,$$ and $$e \eta = - \eta e.$$
We can regard $e$ and $\eta$ as a fundamental pair of Majorana fermions.  
\bigbreak 

Note how the development of the algebra works at this point. We have that $$(e \eta)^2 = -1$$
and so regard this as a natural construction of the square root of minus one in terms of the phase synchronization of the clock that is the iteration of the reentering mark. Once we have the square root of minus one it is natural to introduce another one and call this one $i,$ letting it commute with the other operators. Then we have the $(ie \eta)^2 = +1$ and so we have a triple of Majorana fermions: $$a = e, b = \eta, c = ie \eta$$ and we can construct the quaternions
$$I = ba = \eta e, J = cb = ie,  K = ac = i \eta.$$ With the quaternions in place, we have the braiding 
operators $$A = \frac{1}{\sqrt{2}}(1 + I), B = \frac{1}{\sqrt{2}}(1 + J), C = \frac{1}{\sqrt{2}}(1 + K),$$ 
and can continue as we did above. 
\bigbreak

\section{Laws of Form}
\begin{figure}
     \begin{center}
     \begin{tabular}{c}
     \includegraphics[height=4cm]{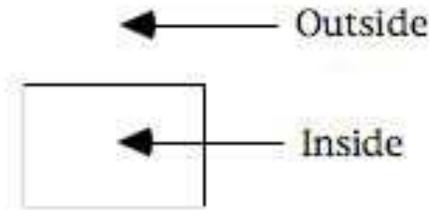}
     \end{tabular}
     \end{center}
     \caption{\bf Inside and Outside}
     \label{outin}
     \end{figure}

This section is a version of a corresponding section in our paper \cite{MLogic}.
Here we discuss a formalism due the G. Spencer-Brown \cite{LOF} that is often called the ``calculus of indications". This calculus is a study of mathematical foundations with a topological notation based on one symbol, the mark:
$$\M{ } \, .$$
This single symbol represents a distinction between its own inside and outside.
As is evident from Fgure~\ref{outin}, the mark is regarded as a shorthand for a rectangle drawn in the plane and dividing the plane into the regions inside and outside the rectangle.   
\bigbreak

     \bigbreak

The reason we introduce this notation is that in the calculus of indications the mark can interact with itself in two possible ways. The resulting formalism becomes a version of Boolean arithmetic, but fundamentally simpler than the usual Boolean arithmetic of $0$ and $1$ with its 
two binary operations and one unary operation (negation). In the calculus of indications one takes a step in the direction of simplicity, and also a step in the direction of physics. The patterns of this
mark and its self-interaction match those of a {\it Majorana fermion}  as discussed in the previous section. A Majorana fermion is a particle that is its own anti-particle.  \cite{ Majorana}. We will later see, in this paper, that by adding braiding to the calculus of indications we arrive at the Fibonacci model, that can in principle support quantum computing.
\bigbreak

In the previous section we described Majorana fermions in terms of their algebra of creation and annihilation operators. Here we describe the particle directly in terms of its interactions. This is part of a general scheme called ``fusion rules" \cite{MooreRead} that can be applied to discrete particle interacations. A fusion rule represents all of the different particle interactions in the form of a set of equations.
The bare bones of the Majorana fermion consist in a particle $P$ such that $P$ can interact with itself to produce a neutral particle $*$ or produce itself $P.$ Thus the possible interactions are 
$$PP \longrightarrow *$$
and 
$$PP \longrightarrow P.$$
This is the bare minimum that we shall need. 
The fusion rule is $$P^2 = 1 + P.$$ This represents the fact that $P$ can interact with itself to produce the neutral particle (represented as $1$ in the fusion rule) or itself (represented by $P$ in the fusion rule). .
\bigbreak

Is there  a {\it linguistic} particle that is its own anti-particle? Certainly we have
$$\sim \sim Q = Q$$ for any proposition $Q$ (in Boolean logic). And so we might 
write 
$$\sim \sim \longrightarrow *$$
where $*$ is a neutral linguistic particle, an identity operator so that
$$*Q = Q$$ for any proposition $Q.$ But in the normal use of negation there is no way that the negation sign combines with itself to produce itself. This appears to ruin the analogy between negation and the 
Majorana fermion. Remarkably, the calculus of indications provides a context in which we can say exactly that a certain logical particle, the mark,  can act as negation {\it and} can interact with itself to produce itself.
\bigbreak

\begin{figure}
     \begin{center}
     \begin{tabular}{c}
     \includegraphics[height=4cm]{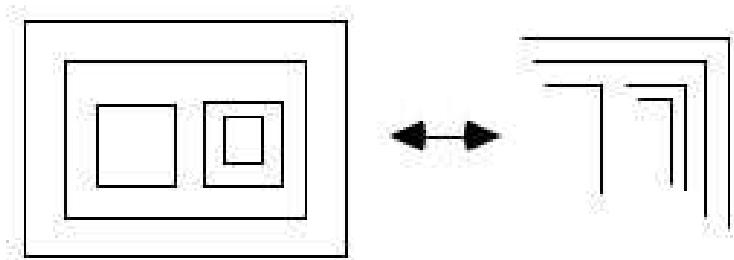}
     \end{tabular}
     \end{center}
     \caption{\bf Boxes and Marks}
     \label{boxmark}
     \end{figure}

In the calculus of indications patterns of non-intersecting marks (i.e. non-intersecting rectangles) are called {\it expressions.} For example in Figure~\ref{boxmark} we see how patterns of boxes correspond to patterns of marks.

     \bigbreak

\begin{figure}
     \begin{center}
     \begin{tabular}{c}
     \includegraphics[height=4cm]{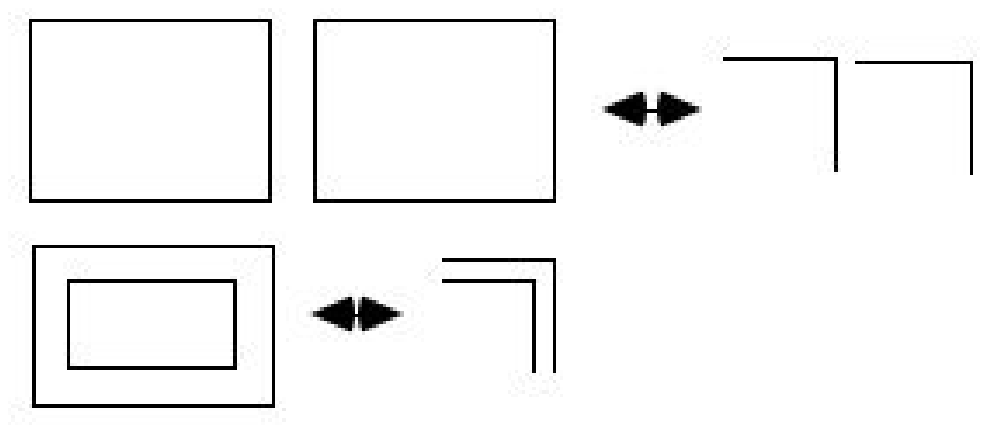}
     \end{tabular}
     \end{center}
     \caption{\bf Translation between Boxes and Marks}
     \label{markbox}
     \end{figure}

In Figure~\ref{boxmark}, we have illustrated both the rectangle and the marked version of the expression.  In an expression you can say definitively of any two marks whether one is or is not inside the other.  The relationship between two marks is either that one is inside the other, or that neither is inside the other.  These two conditions correspond to the two elementary expressions shown in
Figure~\ref{markbox}.

     \bigbreak

The mathematics in Laws of Form begins with two laws of transformation about these two basic expressions. Symbolically, these laws are:
\begin{enumerate}
\item Calling : $$\M{} \, \M{} \, =   \M{}$$  
\item Crossing: $$\M{ \M{ } }  =  \,\,\,\,.$$
\end{enumerate}
The equals sign denotes a replacement step that can be performed on instances of these patterns
(two empty marks that are adjacent or one mark surrounding an empty mark).
In the first of these equations two adjacent marks condense to a single mark, or a single mark expands to form two adjacent marks.  In the second equation  two marks, one inside the other, disappear to form the unmarked state indicated by nothing at all. 
That is, two nested marks can be replaced by an empty word in this formal system.  Alternatively, the unmarked state can be replaced by two nested marks. These equations give rise to a natural calculus, and the mathematics can begin.  For example,  {\it any expression can be reduced uniquely  to either the marked or the unmarked state.}  The he following example illustrates the method:
$$     \M{\M{\M{\M{} \M{}} \M{}} \M{}} \M{}  =   \M{\M{\M{\M{}} \M{}} \M{}}\M{} =   \M{\M{ \M{}} \M{}}\M{} $$
$$ = \M{\M{}}\M{} = \M{} \,\,\,.$$
The general method for reduction is to locate marks that are at the deepest places in the expression
(depth is defined by counting the number of inward crossings of boundaries needed to reach the given mark). Such a deepest mark must be empty and it is either surrounded by another mark, or it is adjacent to an empty mark. In either case a reduction can be performed by either calling or crossing. 
\bigbreak 

Laws of Form begins with the following statement.
``We take as given the idea of a distinction and the idea of an indication, and that it is not possible to make an indication without drawing a distinction. We take therefore the form of distinction for the form."  
Then the author makes the following two statements (laws):
\begin{enumerate}
\item {\it The value of a call made again is the value of the call.}
\item {\it The value of a crossing made again is not the value of the crossing.}
\end{enumerate}
The two symbolic equations above correspond to these statements. First examine the law of calling. It says that the value of a repeated name is the value of the name. In the equation
$$\M{} \, \M{} \, = \M{}$$
one can view either mark as the name of the state indicated by the outside of the other mark.  
In the other equation
$$\M{ \M{ } } = \,\,\,\,.$$
the state indicated by the outside of a mark is the state obtained by crossing from the state indicated on the inside of the mark. Since the marked state is indicated on the inside, the outside must indicate the unmarked state.  The Law of Crossing indicates how opposite forms can fit into one another and vanish into nothing, or how nothing can  produce opposite and distinct forms that fit one another, hand in glove.  The same interpretation yields the equation
$$\M{} \, = \, \M{}$$
where the left-hand side is seen as an instruction to cross from the unmarked state, and the right hand side is seen as an indicator of the marked state. The mark has a double carry of meaning. It can be seen as an operator, transforming the state on its inside to a different state on its outside, and it can be seen as the name of the marked state. That combination of meanings is compatible in this interpretation.  
\bigbreak

From the calculus of indications, one moves to algebra.  Thus 
 $$\M{\M{A}}$$
stands for the two possibilities
  $$\M{\M{\M{}}} \, = \, \M{} \,  \longleftrightarrow  \, A = \M{}$$
$$\M{\M{}} \, = \, \, \, \,  \longleftrightarrow \, A \,  = $$
In all cases we have
$$\M{\M{A}} \, = \, A.$$
 
 By the time we articulate the algebra, the mark can take the role of a unary operator
 $$ A \longrightarrow \M{A}.$$ But it retains its role as an element in the algebra.
Thus begins algebra with respect to this non-numerical arithmetic of forms.  The primary algebra that emerges is a subtle precursor to Boolean algebra.  One can translate back and forth between elementary logic and primary algebra:
\begin{enumerate}
\item $\M{} \longleftrightarrow T$
\item $\M{\M{}} \longleftrightarrow F$
\item $\M{A} \longleftrightarrow \sim A$
\item $AB \longleftrightarrow A \vee B$
\item $\M{\M{A} \M{B}} \longleftrightarrow A \wedge B$
\item $\M{A}B \,\, \longleftrightarrow \,\,A \Rightarrow B$
\end{enumerate}
The calculus of indications and the primary algebra form an efficient system for working with basic symbolic logic.
\bigbreak

By reformulating basic symbolic logic in terms of the calculus of indications, we have a ground in which negation is represented by  the mark {\em and} the mark is also interpreted as a value (a truth value for logic) and these two intepretations are compatible with one another in the formalism. The key to this compatibility is the choice to represent the value ``false"  by a literally unmarked state in the notational plane. With this the empty mark (a mark with nothing on its inside)  can be interpreted as the negation of ``false" and hence represents ``true".
The mark interacts with itself to produce itself (calling) and the mark interacts with itself to produce nothing (crossing). We have expanded the conceptual domain of negation so that it satisfies the mathematical pattern of an abstract Majorana fermion.
\bigbreak

Another way to indicate these two interactions symbolically is to use a box,for the marked state and a blank space for the unmarked state.
Then one has two modes of interaction of a box with itself:
\begin{enumerate}
\item Adjacency: $\fbox{~} ~~ \fbox{~}$
\smallbreak
\noindent and 
\item Nesting: $\fbox{ \fbox{~~} }.$
\end{enumerate}

\noindent With this convention we take the adjacency interaction to yield a single box, and the nesting interaction to produce nothing:

$$\fbox{~} ~~ \fbox{~} = \fbox{~}$$
$$\fbox{ \fbox{~~} } =  $$

\noindent We take the notational opportunity to denote nothing by an asterisk (*). The syntatical rules for operating the asterisk are
Thus the asterisk is a stand-in for no mark at all and it can be erased or placed wherever it is convenient to do so.
Thus $$\fbox{ \fbox{~~} } = *. $$
\bigbreak

At this point the reader can appreciate what has been done if he returns to the usual form of symbolic logic. In that form we that $$\sim \sim X = X$$ for all logical objects (propositions or elements of the logical algebra) $X.$ We can summarize this by writing $$\sim \sim \,\,\,= \,\,\, $$ as a symbolic statement that is outside the logical formalism. Furthermore, one is committed to the interpretation of 
negation as an operator and not as an operand. The calculus of indications provides a formalism where
the mark (the analog of negation in that domain) is both a value and an object, and so can act on itself in more than one way.
\bigbreak

The Majorana particle is its own anti-particle. It is exactly at this point that physics meets logical epistemology. Negation as logical entity is its own anti-particle.  Wittgenstein says (Tractatus \cite{Wittgen} $4.0621$) ``$\cdots$ the sign `$\sim$' corresponds to nothing in reality." And he goes on to say (Tractatus  $5.511$) `` How can all-embracing logic which mirrors the world use such special catches and manipulations? Only because all these are connected into an infinitely fine network, the great mirror." For Wittgenstein in the Tractatus,  the negation sign is part of  the mirror making it possible for thought to reflect reality through combinations of signs. These remarks of Wittgenstein are part of his early picture theory of the relationship of formalism and the world. In our view,  the world and the formalism we use to represent the world are not separate.  The observer and the mark are (formally) identical. A path is opened between logic and physics.
\bigbreak

\begin{figure}
     \begin{center}
     \begin{tabular}{c}
     \includegraphics[height=7cm]{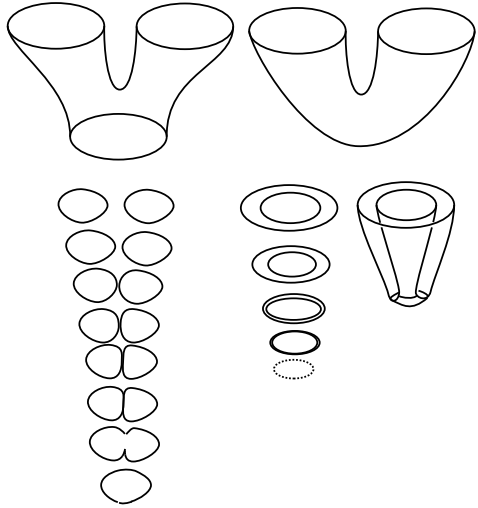}
     \end{tabular}
     \end{center}
     \caption{\bf Calling, Crossing and Cobordism}
     \label{callcross}
     \end{figure}

The visual iconics that create via the boxes of half-boxes of the calculus of indications a model for a logical Majorana fermion can also be seen in terms of cobordisms of surfaces. View Figure~\ref{callcross}. There the boxes have become circles and the interactions of the circles have been displayed as evolutions in an extra dimension, tracing out surfaces in three dimensions. The condensation of two circles to one is a simple cobordism betweem two circles and a single circle. The cancellation of two circles that are concentric can be seen as the right-hand lower cobordism in this figure with a level having a continuum of critical points where the two circles cancel. A simpler cobordism is illustrated above on the right where the two circles are not concentric, but nevertheless are cobordant to the empty circle. Another way of putting this is that two topological closed strings can interact by cobordism to produce a single string or to cancel one another. Thus a simple circle can be a topological model for a Majorana fermion.  

     \bigbreak

In \cite{MLogic,AnyonicTop} we detail how the Fibonacci model for anyonic quantum computing can be constructed by using a version of
the two-stranded  bracket polynomial and a generalization of Penrose spin networks. This is a fragment of the Temperly-Lieb recoupling theory \cite{KL}.  
\bigbreak

\subsection{The Square Root of Minus One Revisited}
 So far we have seen that the mark can represent the fusion rules for a Majorana fermion since it can interact with itself to produce either itself or nothing. But we have not yet seen the anti-commuting fermion algebra emerge from this context of making a distinction. Remarkably, this algebra does emerge when one looks at the mark recursively.  
 \bigbreak
 
\begin{figure}
     \begin{center}
     \begin{tabular}{c}
     \includegraphics[width=6cm]{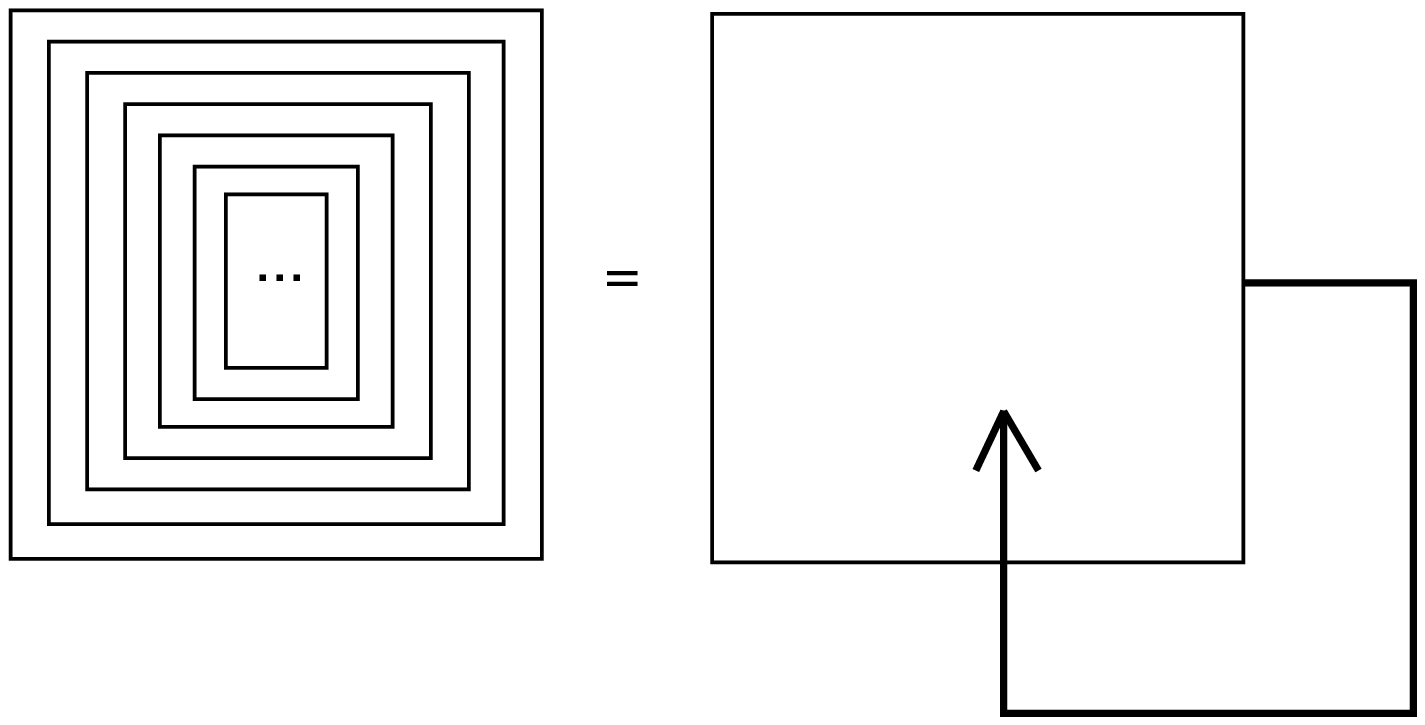}
     \end{tabular}
     \caption{\bf}
     \label{fix}
\end{center}
\end{figure}

 Consider the transformation $$F(X) = \M{X}.$$
 If we iterate it  and take the limit we find 
 $$G = F(F(F(F( \cdots )))) = \M{\M{\M{\M{ ... }}}}$$ 
 an infinite nest of marks satisfying the equation
 $$G = \M{G}.$$
 With $G = F(G),$ I say that $G$ is an {\em eigenform} for the transformation $F.$
 See Figure~\ref{fix} for an illustration of this nesting with boxes and an arrow that points inside the reentering mark to indicate its appearance inside itself. If one thinks of the mark itself as a Boolean logical value, then extending the language to include the reentering mark $G$ goes beyond the boolean. We will not detail here how this extension can be related to non-standard logics, but refer the reader to \cite{KL}. Taken at face value the reentering mark cannot be just marked or just unmarked, for by its very definition, if it is marked then it is unmarked and if it is unmarked then it is marked. In this sense the reentering mark has the form of a self-contradicting paradox. There is no paradox since we do not have to permanently assign it to either value. The simplest interpretation of the reentering mark is that it is temporal and that it represents an oscillation between markedness and unmarkedness.  In numerical terms it is a discrete dynamical system oscillating between $+1$ (marked) and $-1$ (not marked).
 \bigbreak
 
 With the reentering mark in mind consider now the transformation on real numbers given by 
 $$T(x) = -1/x.$$ This has the fixed points $i$ and $-i$, the complex numbers whose squares are 
 negative unity. But lets take a point of view more directly associated with the analogy of the recursive mark. Begin by starting with a simple periodic process that is associated directly 
with the classical attempt to solve for $i$ as a solution to a quadratic equation. We take the point
of view that solving $x^2 = ax + b$ is the same (when $x \ne 0$) as solving 
$$ x = a + b/x,$$ and hence is a matter of finding a fixed point. In the case of $i$ we have
$$x^2 = -1$$ and so desire a fixed point $$x = -1/x.$$ There are no real numbers that are fixed points for this operator and so we consider the oscillatory process generated by
$$T(x) = -1/x. $$
The fixed point would satisfy
$$i = -1/i$$
and multiplying, we get that 
$$ii = -1.$$
On the other hand the iteration of $T$ yields
$$ 1, T(1) = -1 , T(T(1)) = +1 , T(T(T(1))) = -1, +1,-1,+1,-1, \cdots .$$
The square root of minus one is a perfect example of an eigenform that occurs in a new and wider domain than the original context in which its recursive process arose. The process has no fixed point in the original domain. At this point we enter once again the domain of iterants and particularly the discussion of Section 6 where we see the square root of minus one as a clock.
\bigbreak

There is one more comment that is appropriate for this section. Recall that a pair of Majorana fermions can be assembled to form a single standard fermion.  In our case we have the spatial and temporal iterant components  $e =[1,-1]$ and $\eta$ with $e\eta = -\eta e.$
We can regard $e$ and $\eta$ as a fundamental pair of Majorana fermions. This is a formal correspondence, but it is striking how this Marjorana fermion algebra emerges from an analysis of the 
recursive nature of the reentering mark, while the fusion algebra for the Majorana fermion emerges from the distinctive properties of the mark itself. We see how the seeds of the fermion algebra live  in this extended logical context.
\bigbreak 
 
 The corresponding standard fermion annihilation and creation operators are then given by the formulas below.
$$\psi = (e + i\eta)/2$$ and
$$\psi^{\dagger} = (e - i\eta)/2.$$
Since $e$ represents a spatial view of the basic discrete oscillation and $\eta$ is the time-shift operator for this oscillation it is of interest to note that the standard fermion built by these two can be regarded as a quantum of spacetime, retrieved from the way that we decomposed the process into space and time. Since all this is initially built in relation to extending the Boolean logic of the mark to a non-boolean recursive context, there is further analysis needed of the relation of the physics and the logic. We have only begun the analysis here. The crux of the matter is that two dimensional physics depends upon a plane space in which a simple closed curve makes a distinction between inside and outside in order for the braiding and phases to be significant. This same property of distinction in the plane is what gives a plane space the linguistic power to represent language and logic. This correspondence in not an accident and deserves further study!
\bigbreak

\section{The Dirac Equation and Majorana Fermions}
 We now construct the Dirac equation. This may sound circular, in that the fermions arise from solving the Dirac equation, but in fact the algebra underlying this equation has the same properties as the  creation and  annihilation algebra for fermions, so it is by way of this algebra that we will come to the Dirac equation. If the speed of light is equal to $1$ (by convention), then energy $E$, momentum $p$ and mass $m$ are
related by the (Einstein) equation $$E^2 = p^2 + m^2.$$ Dirac constructed his equation by looking
for an algebraic square root of $ p^2 + m^2$ so that he could have a linear operator for $E$ that would take the same role as the Hamiltonian in the Schroedinger equation. We will get to this operator by first taking the case where $p$ is a scalar (we use one dimension of space and one dimension of time.).
Let $E = \alpha p + \beta m$ where $\alpha$ and $\beta$ are elements of a a possibly non-commutative,
associative algebra. Then $$E^2 = \alpha^2 p^2 + \beta^2 m^2 + pm(\alpha \beta + \beta \alpha).$$
Hence we will satisfiy $E^2 = p^2 +m^2$ if $\alpha^2 = \beta^2 = 1$ and
 $\alpha \beta + \beta \alpha = 0.$ This is our familiar Clifford algebra pattern and we can use the iterant algebra generated by $e$ and $\eta$ if we wish. Then, because the quantum operator for momentum is $-i \partial/\partial x$ and the operator for energy is $i\partial/\partial t,$ we have the Dirac equation $$i\partial \psi /\partial t = -i \alpha \partial \psi /\partial x + \beta m \psi.$$
 Let $${\cal O} = i\partial /\partial t + i \alpha \partial /\partial x - \beta m $$ so that the Dirac equation 
 takes the form $${\cal O} \psi(x,t) = 0.$$ Now note that 
 $${\cal O} e^{i(px - Et)} = (E - \alpha p - \beta m) e^{i(px - Et)}.$$
 We let $$\Delta = (E - \alpha p - \beta m)$$ and let 
 $$ U = \Delta \beta \alpha = (E - \alpha p - \beta m)\beta \alpha = \beta \alpha E + \beta p - \alpha m,$$
 then $$U^2 = -E^2 + p^2 + m^2 = 0.$$ This nilpotent element leads to a (plane wave) solution to the 
 Dirac equation as follows:  We have shown that $${\cal O} \psi = \Delta \psi$$ for 
 $\psi =  e^{i(px - Et)}.$  It then follows that 
 $${\cal O}(\beta \alpha \Delta \beta \alpha \psi) = \Delta \beta \alpha \Delta \beta \alpha  \psi =
 U^2 \psi = 0,$$
  from which it follows that $$\psi = \beta \alpha U e^{i(px - Et)}$$ is a (plane wave) solution to the Dirac equation.
\bigbreak

In fact, this calculation suggests that we should multiply the operator ${\cal O}$ by $\beta \alpha$ on the right, obtaining the operator 
$${\cal D} = {\cal O} \beta \alpha =  i\beta \alpha \partial /\partial t + i \beta \partial /\partial x - \alpha m, $$
and the equivalent Dirac equation $${\cal D}\psi = 0.$$ In fact for the specific $\psi$ above we will now 
have ${\cal D} (U  e^{i(px - Et)} ) = U^2  e^{i(px - Et)} = 0.$ This idea of reconfiguring the Dirac equation
in relation to nilpotent algebra elements $U$ is due to Peter Rowlands \cite{Rowlands}.  Rowlands does this in the context of quaternion algebra.
Note that the solution to the Dirac equation that we have found is expressed in Clifford algebra or iterant algebra form. It can be articulated into specific vector solutions by using an iterant or matrix representation of the algebra.
\bigbreak

We see that $ U = \beta \alpha E + \beta p - \alpha m$ with $U^2 = 0$ is really the essence of  this plane wave solution to the Dirac equation. This means that a natural non-commutative algebra arises directly 
and can be regarded as the essential information in a Fermion. It is natural to compare this algebra structure with algebra of creation and annihilation operators that occur in quantum field theory.
to this end, let $$U^{\dagger} = \alpha \beta E + \alpha p - \beta m.$$ Here we regard $U^{\dagger}$
as a formal counterpart to complex conjugation, since in the split quaternion algebra we have not yet constructed commuting square roots of negative one. We then find that with 
$$ A = U + U^{\dagger} = (\alpha + \beta)(p-m)$$ and 
$$ B = U - U^{\dagger} = 2 \beta \alpha E + (\beta - \alpha)(p-m)$$ that 
$$[\frac{A}{\sqrt{2}(p-m)}]^{2} = 1$$ and 
$$[\frac{iB}{\sqrt{2}(p+m)}]^{2} = 1,$$ with $i$ a commuting square root of negative one, giving the 
underlying Majorana Fermion operators for our Dirac Fermion. The operators $U$ and $U^{\dagger}$
satisfy the usual commutation relations for the annihilation and creation operators for a Fermion.
\bigbreak

It is worth noting how the Pythgorean relationship $E^2 = p^2 + m^2$ interacts here with the Clifford algebra of $\alpha$ and $\beta.$ We have
$$ U^{\dagger} =p \alpha + m \beta +  \alpha \beta E$$ 
$$ U =p \beta + m \alpha + \beta \alpha E$$ with
$$(U^{\dagger})^2 = U^2 = 0,$$
$$U + U^{\dagger} = (p + m)(\alpha + \beta),$$
$$U - U^{\dagger} = (p - m)(\alpha - \beta) + 2 E \alpha \beta.$$
This implies that $$(U + U^{\dagger})^2 = 2(p +m)^2$$
$$(U - U^{\dagger})^2 = 2(p-m)^2 - 4E^2 = 2[p^2 + m^2 - 2pm - 2p^2 - 2m^2] = -2(p+m)^2.$$
From this we easily deduce that 
$$U U^{\dagger} + U^{\dagger} U = 2 (p + m)^2,$$ and this can be normalized to equal $1.$

\subsection{Another version of $U$ and $U^{\dagger}$}
We start with $\psi =  e^{i(px - Et)}$ and the operators
$$\hat{E} = i \partial/\partial t$$ and 
$$\hat{p} = -i \partial/\partial x$$  so that 
$$\hat{E}\psi = E \psi$$ and
$$\hat{p}\psi = p \psi.$$
The Dirac operator is
$${\cal O} = \hat{E} - \alpha \hat{p} - \beta m$$
and the modified Dirac operator is
$${\cal D} ={\cal O} \beta \alpha =  \beta \alpha  \hat{E} + \beta \hat{p} - \alpha m,$$ so that
$$ {\cal D}\psi = (\beta \alpha E + \beta p - \alpha m)\psi = U \psi.$$
If we let $$\tilde{\psi} =  e^{i(px +Et)}$$ (reversing time), then we have
$$ {\cal D}\tilde{\psi} = (-\beta \alpha E + \beta p - \alpha m)\psi = U^{\dagger} \tilde{\psi},$$
giving a definition of $U^{\dagger}$ corresponding to the anti-particle for $U\psi.$
\bigbreak

We have
$$U =  \beta \alpha E + \beta p - \alpha m$$
and
$$U^{\dagger} =  - \beta \alpha E + \beta p - \alpha m$$
Note that here we have 
$$(U + U^{\dagger})^2 =  (2 \beta p + \alpha m)^2 =  4 (p^2 + m^2 )= 4 E^2 ,$$ and
$$(U - U^{\dagger})^2 = - ( 2 \beta \alpha E)^2 = - 4 E^2 .$$
We have that $$U^{2} = (U^{\dagger})^{2} = 0 $$ and $$U U^{\dagger} + U^{\dagger} U = 4 E^{2}.$$
Thus we have a direct appearance of the Fermion algebra corresponding to the Fermion plane wave solutions to the Dirac equation. Furthermore, the decomposition of $U$and $U^{\dagger}$ into the
corresponding Majorana Fermion operators corresponds to $E^2 = p^2 + m^2 .$
Normalizing by dividing by $2 E$ we have
$$A =( \beta p + \alpha m)/E $$ and 
$$B = i \beta \alpha.$$ so that
$$A^2 = B^2 = 1$$ and $$AB + BA = 0.$$ then
$$U = (A + Bi)E$$ and $$U^{\dagger} = (A - Bi)E, $$ showing how the Fermion operators are expressed in terms of the simpler Clifford algebra of Majorana operators (split quaternions once again).

\bigbreak

\subsection{Writing in the Full Dirac Algebra}
We have written the Dirac equation so far in one dimension of space and one dimension of time.
We give here a way to boost the formalism directly to three dimensions of space. We take an independent Clifford algebra generated by $\sigma_{1}, \sigma_{2}, \sigma_{3}$ with
$\sigma_{i}^{2} = 1$ for $i=1,2,3$ and $\sigma_{i}\sigma_{j} = - \sigma_{j}\sigma_{i}$ for 
$i \ne j.$ Now assume that $\alpha$ and $\beta$ as we have used them above generate an independent Clifford algebra that commutes with the algebra of the $\sigma_{i}.$ Replace
the scalar momentum $p$ by a $3$-vector momentum $p = (p_1 , p_2 , p_3 )$ and let 
$p \bullet \sigma = p_{1} \sigma_{1} +  p_{2} \sigma_{2} +  p_{3} \sigma_{3}.$ We replace
$\partial / \partial x$ with $\nabla  = (\partial /  \partial x_{1} , \partial / \partial x_{2}, \partial / \partial x_{2} )$
and $\partial p / \partial x$ with $\nabla \bullet p.$
\bigbreak

We then have the following form of the Dirac equation.
$$i\partial \psi /\partial t = -i \alpha \nabla \bullet \sigma  \psi  + \beta m \psi.$$
 Let $${\cal O} = i\partial /\partial t + i \alpha \nabla \bullet \sigma  - \beta m $$ so that the Dirac equation 
 takes the form $${\cal O} \psi(x,t) = 0.$$  In analogy to our previous discussion we let 
 $$\psi(x,t) =  e^{i(p \bullet x - Et)}$$ and construct solutions by first applying the Dirac operator to this $\psi.$ The two Clifford algebras interact to generalize directly the nilpotent solutions and Fermion algebra that we have detailed for one spatial dimension to this three dimensional case. To this purpose the modified Dirac operator is
$$ {\cal D} = i\beta \alpha \partial/\partial t + \beta \nabla \bullet \sigma - \alpha m.$$
And we have that $${\cal D}\psi = U \psi$$ where
$$U = \beta \alpha E + \beta p \bullet \sigma - \alpha m.$$
We have that $U^{2}= 0$ and  $U \psi$ is a solution to the modified Dirac Equation, just as before.
And just as before, we can articulate the structure of the Fermion operators and locate the corresponding Majorana Fermion operators. We leave these details to the reader.
 \bigbreak
 
 \subsection{Majorana Fermions at Last}
There is more to do. We will end with a brief discussion making  Dirac algebra distinct from the one generated by $\alpha, \beta, \sigma_1 , \sigma_2 , \sigma_3$ to obtain an equation that can have real solutions. This was the strategy that Majorana \cite{Majorana} followed to construct his Majorana Fermions. A real equation can have solutions that are invariant under complex conjugation and so can correspond to particles that are their own anti-particles. We will describe this Majorana algebra in terms of the split quaternions $\epsilon$ and $\eta.$ For convenience we use the matrix representation given below. The reader of this paper can substitute the corresponding iterants.

$$ \epsilon = \left(\begin{array}{cc}
			-1&0\\
			 0&1
			\end{array}\right),
			 \eta = \left(\begin{array}{cc}
			0&1\\
			1&0
			\end{array}\right).$$
Let $\hat{\epsilon}$ and $\hat{\eta}$ generate another, independent algebra of 
split quaternions, commuting with the first algebra generated by $\epsilon$ and $\eta.$
Then a totally real Majorana Dirac equation can be written as follows:
$$(\partial/\partial t + \hat{\eta} \eta \partial/\partial x + \epsilon \partial/\partial y + \hat{\epsilon} \eta \partial/\partial z - \hat{\epsilon} \hat{\eta} \eta m) \psi = 0.$$
\bigbreak

\noindent To see that this is a correct  Dirac equation, note that
$$\hat{E} = \alpha_{x} \hat{p_{x}} +  \alpha_{y} \hat{p_{y}} +  \alpha_{z} \hat{p_{z}} + \beta m$$
(Here the ``hats'' denote the quantum differential operators corresponding to the energy and momentum.)
will satisfy $$\hat{E}^{2} =  \hat{p_{x}}^{2} + \hat{p_{y}}^{2} + \hat{p_{z}}^{2} + m^{2}$$ if the algebra
generated by $\alpha_{x}, \alpha_{y}, \alpha_{z}, \beta$ has each generator of square one and each distinct pair of generators anti-commuting. From there we obtain the general Dirac equation by replacing
$\hat{E}$ by $i\partial/\partial t$, and $\hat{p_{x}}$ with $-i\partial/\partial x$ (and same for $y,z$).
$$(i\partial/\partial t +i\alpha_{x}\partial/\partial x +i\alpha_{y} \partial/\partial y +i\alpha_{z} \partial/\partial y - \beta m) \psi = 0.$$ 
This is equivalent to
$$(\partial/\partial t  +\alpha_{x}\partial/\partial x  + \alpha_{y} \partial/\partial y + \alpha_{z} \partial/\partial y +i \beta m) \psi = 0.$$
Thus, here we take $$\alpha_{x} = \hat{\eta} \eta, \alpha_{y} =  \epsilon, \alpha_{z} = \hat{\epsilon} \eta ,
\beta = i\hat{\epsilon} \hat{\eta} \eta,$$ and observe that these elements satisfy the requirements for the Dirac algebra. Note how we have a significant interaction between the commuting square root of minus one ($i$) and the element $\hat{\epsilon} \hat{\eta}$ of square minus one in the split quaternions. This brings us back to our original considerations about the source of the square root of minus one. Both viewpoints combine in the element $\beta = i \hat{\epsilon} \hat{\eta} \eta$ that makes this Majorana algebra work. Since the algebra appearing in the Majorana Dirac operator is constructed entirely from two commuting copies of the split quaternions, there is no appearance of the complex numbers, and when written out in $2 \times 2$ matrices we obtain coupled real differential equations to be solved. Clearly this ending is actually a beginning of a new study of Majorana Fermions. That will begin in a sequel to the present paper.
\bigbreak


\begin{thebibliography}{99}


\bibitem{LOF}
G. Spencer--Brown, ``Laws of Form," George Allen and Unwin Ltd. London (1969).


 \bibitem{SS}
Kauffman, L. [1985], Sign and Space,  In Religious Experience and Scientific Paradigms. Proceedings of
the 1982 IASWR Conference, Stony Brook, New York: Institute of Advanced Study of
World Religions, (1985), 118-164.

\bibitem{SRF}
Kauffman, L. [1987], Self-reference and recursive forms,  Journal of Social and Biological Structures 
(1987), 53-72.

\bibitem{SRCD}
Kauffman, L. [1987], Special relativity and a calculus of distinctions.  Proceedings of the 9th Annual
Intl. Meeting of ANPA, Cambridge, England (1987).  Pub. by ANPA West, pp.
290-311.

\bibitem{IML}
Kauffman, L. [1987], Imaginary values in mathematical logic.  Proceedings of the Seventeenth
International Conference on Multiple Valued Logic, May 26-28 (1987), Boston MA,
IEEE Computer Society Press, 282-289.


\bibitem{KL} 
Kauffman, L. H.,  Knot Logic,  In {\it Knots and Applications}  ed. by L. Kauffman, World Scientific Pub. Co., (1994), pp. 1-110.

\bibitem{Majorana} E. Majorana, A symmetric theory of electrons and positrons, 
I Nuovo Cimento,{\bf 14} (1937), pp. 171-184.

\bibitem{MooreRead}
G. Moore and N. Read, Noabelions in the fractional quantum Hall effect, 
{\em Nuclear Physics B360} (1991), 362 - 396.


\bibitem{Kouwenhouven} 
V. Mourik,K. Zuo, S. M. Frolov, S. R. Plissard, E.P.A.M. Bakkers, L.P. Kouwenhuven, Signatures of Majorana fermions in hybred superconductor-semiconductor devices,
arXiv: 1204.2792.

\bibitem{BL}
Kauffman, Louis H. [2002], Biologic. {\em AMS Contemporary Mathematics Series},
Vol. 304, (2002), pp. 313 - 340.

\bibitem{Kauff:KP} Kauffman,Louis H.[1991,1994,2001,2012], {\em Knots and Physics,}
World Scientific Pub.

\bibitem{KL}
L.H. Kauffman, {\em Temperley-Lieb Recoupling Theory and Invariants of Three-Manifolds},
Princeton University Press, Annals Studies {\bf 114} (1994). 

\bibitem{Para}
Kauffman, Louis H. [2002], Time imaginary value, paradox sign and space,
in {\em Computing Anticipatory Systems, CASYS -
Fifth International Conference}, Liege, Belgium (2001) ed. by Daniel Dubois,
AIP Conference Proceedings Volume 627 (2002).


\bibitem{KN:QEM} Kauffman,Louis H. and Noyes,H. Pierre [1996], Discrete
Physics and the Derivation of Electromagnetism from the formalism of
Quantum Mechanics, {\em Proc. of the Royal Soc. Lond. A}, {\bf 452}, pp.
81-95. 

\bibitem{KN:Dirac} Kauffman,Louis H. and Noyes,H. Pierre [1996], Discrete
Physics and the Dirac Equation, {\em Physics Letters A}, 218 ,pp.
139-146. 

\bibitem{NonCom} Kauffman, Louis H. [1998], Noncommutativity and discrete
physics, {\em Physica D } 120 (1998), 125-138.

\bibitem{ST} Kauffman, Louis H. [1998], Space and time in discrete physics, 
{\em Intl. J. Gen. Syst.} Vol. 27, Nos. 1-3, 241-273.

\bibitem{Aspects}
Kauffman, Louis H. [1999], A non-commutative approach to discrete physics, 
in {\em Aspects II - Proceedings of ANPA 20}, 215-238.

\bibitem{Boundaries}
Kauffman, Louis H. [2003], Non-commutative calculus and discrete physics,
in {\em Boundaries- Scientific Aspects of ANPA 24}, 73-128.

\bibitem{NCW}
Kauffman, Louis H. [2004], Non-commutative worlds, {\em New Journal of Physics 6}, 2-46.

\bibitem{Noyes}
Kauffman, Louis H., Non-commutative worlds and classical constraints. in ``Scientific Essays in Honor of Pierre Noyes on the Occasion of His 90-th Birthday", edited by John Amson and Louis H. Kaufman, World Scientific Pub. Co. (2013), pp. 169-210.

\bibitem{Fauser}
Louis H. Kauffman, Differential geometry in non-commutative worlds, in ``Quantum Gravity - 
Mathematical Models and Experimental Bounds", edited by B. Fauser, J. Tolksdorf and E. Zeidler,
Birkhauser (2007), pp. 61 - 75.

\bibitem{MLogic}
Kauffman, Louis H. [2012],  Knot Logic and  Topological Quantum Computing with Majorana Fermions,
(to appear).

\bibitem{AnyonicTop}
Kauffman, Louis H.; Lomonaco, Samuel J., Jr. $q$-deformed spin networks, knot polynomials and anyonic topological quantum computation. J. Knot Theory Ramifications 16 (2007), no. 3, 267--332. 

\bibitem{Littlewood}
D. E. Littlewood, ``The Skeleton Key of Mathematics", Harper Torchbook Edition (1960).

\bibitem{Rowlands} Peter Rowlands, ``Zero to Infinity - The Foundations of Physics'',  Series on Knots
and Everything - Volume 41, World Scientific Publishing Co., 2007.

\bibitem{Wittgen} L. Wittgenstein, ``Tractatus Logico - Philosophicus", New York: Harcourt, Brace and Company, Inc., London: Kegan Paul, Trench, Trubner and Co. Ltd. (1922).

\end{thebibliography}
\end{document}